\documentclass[a4paper,11pt]{article}
\pdfoutput=1
\usepackage{jcappub}
\usepackage{soul}
\usepackage{xcolor}
\usepackage{cleveref} 
    \crefname{equation}{}{}
    \crefname{align}{}{}
    \crefname{figure}{}{}    
    \crefname{table}{}{}
    \crefname{section}{}{}   
    \crefname{appendix}{}{}
    \crefname{footnote}{}{}
    
    \crefalias{subequation}{equation}

\usepackage{orcidlink}

\def\nc{\newcommand}

\def\hrr{h_{rr}}

\def\htt{h_{\theta\theta}}

\definecolor{orcid-green}{RGB} {166, 206, 57}

\def\at{a_\theta}
\def\rat{ra_{\theta}}
\def\ratr{(ra_\theta)_{,r}}

\def\ratp{(ra_\theta)'}
\def\ratpp{(ra_{\theta})''}
\def\drat{r \dot a_\theta}
\def\dratp{(r\dot a_\theta)'}
\def\dat{{\dot a}_\theta}
\def\ddat{{\ddot a}_\theta}
\def\atp{a_\theta'}
\def\atpp{a_\theta''}

\def\ar{a_r}
\def\dar{{\dot a}_r}

\def\ddar{{\ddot a}_r}
\def\ddat{{\ddot a}_\theta}
\def\arp{a_r'}

\def\KRr{{K^r}_r}
\def\KTt{{K^\theta}_\theta}

\def\KTtr{{K^\theta}_{\theta,r}}

\def\equals{\;\,=\;\,}
\def\isequiv{\;\,\equiv\;\,}

\def\DE{\scriptscriptstyle \rm DE}
\def\DM{\scriptscriptstyle \rm DM}
 
\nc{\lsim}{\;\raisebox{-.3em}{$\stackrel{\displaystyle <}{\sim}$}\;}
\nc{\gsim}{\;\raisebox{-.3em}{$\stackrel{\displaystyle >}{\sim}$}\;}

\definecolor{pink}{rgb}{0.8, 0.0, 0.8}
\usepackage[titletoc,toc,page]{appendix}

\subheader{CERN-TH-2026-244}
\title{LazyTB: Extended LTB-solutions with Multiple Interacting Fluids}

\author[a,b,c]{Kimmo Kainulainen\orcidlink{0000-0002-1165-9375},}
\author[d]{Enrico D.~Schiappacasse\orcidlink{0000-0002-6136-1358},}
\author[a]{Linda Tenhu\orcidlink{0000-0002-7746-8512},}
\author[a,b]{Olli Väisänen\orcidlink{0000-0002-5089-8512}}

\affiliation[a]{Department of Physics, PL 35 (YFL), 40014 University of Jyv\"askyl\"a, Finland}

\affiliation[b]{Helsinki Institute of Physics, PL 64, 00014 University of Helsinki, Finland}

\affiliation[c]{Theoretical Physics Department, CERN, CH-1211 Genève, Switzerland}

\affiliation[d]{Facultad de Ingenier\'ia, Universidad San Sebasti\'an, Bellavista 7, Santiago 8420524, Chile}

\emailAdd{kimmo.kainulainen@jyu.fi}
\emailAdd{enrico.schiappacasse@uss.cl}
\emailAdd{astrotenhu@gmail.com}
\emailAdd{olli.j.r.vaisanen@jyu.fi}

\abstract{We present a low entry barrier numerical code LazyTB that solves the Einstein equations in the ADM-formulation for a general spherically symmetric problem with an arbitrary number of fluids with user defined equations of state and optional interactions between the fluids mediated by drag-vectors. We present the ADM equations in terms of carefully chosen dimensionless
variables that avoid some numerical instabilities by analytic cancellation of potentially diverging terms. We demonstrate the code by applying it to a number of examples including the standard LTB model, a model with radiation and dark matter, a polytropic dark energy model with a non-vanishing sound speed and a model with coupled interacting dark matter and dark energy. We also implement a suite of diagnostic tools, including ray-tracing algorithms, that allow extracting physical observables such as luminosity distances for observers at arbitrary locations from the simulations. The code is built to be easy to extend also beyond the late-universe applications considered in the examples of this paper. LazyTB is available at
https://github.com/ollijvaisanen/LazyTB.}


\arxivnumber{astro-ph/2020xxx}

\begin{document}
\maketitle

%
%

%
\section{Introduction}
%

The current cosmological model of $\Lambda$CDM has been remarkably successful in explaining a wide variety of observations. However, discrepancies in the model, such as the Hubble tension \cite{Riess:2024ohe,Riess:2021jrx,Pesce:2020xfe, Riess:2019cxk,Riess:2016jrr} and the recent DESI measurements \cite{DESI:2025zgx, DESI:2025zpo}, have begun to emerge in the past decade, which has led to a breadth of literature on various modified models and proposed solutions including, for example, early dark energy models~\cite{Chaussidon:2025npr,Poulin:2023lkg,Herold:2022iib,Poulin:2018cxd}, interacting dark sectors models~\cite{Petri:2025swg,Bagherian:2024obh, Chudaykin:2020igl,  DiValentino:2017iww}, modified gravity and alternative cosmologies~\cite{Clifton:2024mdy, Tiwari:2023jle, 2022InJPh..96.1289S, 2020arXiv201112146G}, and systematic errors~\cite{Banik:2025dlo, Sharma:2025ucg}. One approach to reconciling these discrepancies is to treat them as a sign of new physics, inviting us to study, for instance, various models of dark matter and dark energy. Another approach would be to search for solutions within conventional physics, such as the large-scale structure of the universe or the backreaction of nonlinear physics on smaller scales. Although each of these approaches has been well-studied in the literature, their combined effects are rarely considered due to the cumbersome nature of actually implementing them in practice.

There are some available numerical codes in the literature within the framework of spherically symmetric spacetimes. For example, the EasyLTB is a Fortran code designed to test Lemaitre-Tolman-Bondi (LTB) cosmological models against observational data, such as those from Type Ia Supernovae, Cosmic Microwave Background, and Baryon Acoustic Oscillations~\cite{Garcia-Bellido:2008vdn}. The code uses arbitrary functions for the radial dependence of the matter density and the Hubble expansion rate, implying a single fluid component. The EasyLTB code was modified by Ref.~\cite{2012ApJ...748..111W} to include a non-zero curvature. A collection of numerical codes for general relativistic cosmology, written in Python and C++\, was released in~\cite{WinNT}, and in~\cite{Bull:2011wi} they were used in spherically-symmetric, dust-only radially-inhomogeneous LTB models to derive constraints from observations of the kinematic Sunyaev-Zel'dovich (kSZ) effect. More recently, the monteLLTB code~\cite{Camarena_2021} has been released to provide a way of sampling $\Lambda$LTB-models using Monte Carlo methods.

Due to the reduced utility of specialized LTB codes in more generic situations and the complexity of more general-purpose GR-codes, we believe there is a interest for a solver with a low entry barrier for cosmologists not specializing in numerical relativity. For this purpose, we present LazyTB, a publicly available code for computing cosmological observables in spherically symmetric spacetimes with arbitrary energy components. The code consists of a solver for the fully nonlinear metric, diagnostic tools for its numerical error, and a solver for observables such as angular distances along light-ray trajectories for an observer at an arbitrary location in the spacetime. To this end, we write the spherical ADM equations alongside the evolution equation for general fluids in a form that avoids numerical instabilities. The code allows for an arbitrary number of fluids with user-defined equations of state as well as optional fluid interactions described by user-defined drag-vectors. It is also built for ease of extension beyond the setups and observables considered in this paper.

This paper is organized as follows: First, we present the derivation of the system of equations for the metric components and a collection of perfect fluids starting from the ADM formalism. Our discussion is brief but self-contained. We also discuss the definition of the initial values and the boundary conditions, and derive equations for computing physical observables along light trajectories. In section three we validate the code and demonstrate its use applying it to several examples, including cosmologies with pure dust, evolving dark energy and interacting dark matter and dark energy. More details of the numerical implementation, including recasting equations in dimensionless variables, and the GitHub repository are provided in the appendix and the conclusions. 

%
\section{Spherically symmetric models}
%

In this section, we begin by presenting concise, first-order evolution equations for a spherically symmetric metric using the ADM formalism ~\cite{Arnowitt:1962hi, Baumgarte:1998te}. In particular, we introduce a modified set of variables which improves the numerical stability of the system, and then derive equations of motion for the fluid densities and velocities in these new variables. Following this, we will discuss the appropriate initial and boundary conditions for a spherically symmetric perfect multicomponent fluid embedded in an FLRW background. Finally, we will discuss how to calculate observables by solving the geodesic equations and the Sachs optical equations in the solution metric.

%
\subsection{General ADM evolution equations}
%

The most general metric can be decomposed as~\cite{Thiemann:2007zz,Baumgarte:2010ndz}:
\begin{equation}
{\rm d}s^2 = - \alpha^2 {\rm d}t^2 
             + h_{ij}({\rm d}x^i + \beta^i{\rm d}t)({\rm d}x^j + \beta^j{\rm d}t)
\label{eq:ds2}             
\end{equation}
The ADM evolution equations are written for the lapse function $\alpha$, shift vector $\beta^i$ and the 3-metric $h_{ij}$ appearing in the metric~\cref{eq:ds2}, and for an additional variable, the extrinsic curvature tensor $K_{ij}$, as follows~\cite{Arnowitt:1962hi,Baumgarte:1998te}:
\begin{equation}
{\rm d}_t h_{ij} = - 2 \alpha K_{ij}
\label{eq:BS1}
\end{equation}
and
\begin{equation}
{\rm d}_t K_{ij} = - D_iD_j\alpha + \alpha ( R_{ij} - 2K_{il}{K^l}_j + KK_{ij}-\kappa M_{ij}),
\label{eq:BS2}
\end{equation}
where ${\rm d}_t = \partial_t - {\cal L}_\beta$ and ${\cal L}_\beta$ denotes the Lie derivative with respect to $\beta^i$ and $D_i$ is the covariant derivative with respect to the three metric. Moreover, the gravitational coupling is defined as $\kappa \equiv 8\pi G_N$ and finally $K \equiv h^{ij}K_{ij}$ is the trace of the extrinsic curvature. In addition the system is bound by two constraints: the {\em Hamiltonian constraint}:
\begin{equation}
R - K_{ij}K^{ij} + K^2 = 2 \kappa \hat \rho
\label{eq:C1}
\end{equation}
and the {\em momentum constraint}:
\begin{equation}
D_l{K^l}_i  - D_iK = \kappa S_i\,.
\label{eq:C2}
\end{equation}
Here $R = h^{ij}R_{ij}$, where $R_{ij}$ is the three dimensional Ricci tensor. The source terms $\hat \rho$, $S_{i}$ and 
\begin{equation}
M_{ij} \equiv S_{ij} + \frac{1}{2}h_{ij} (\hat \rho - h^{kl}S_{kl} )
\end{equation}
appearing in equations~\cref{eq:BS2,eq:C2} are related to the energy-momentum tensor through the following projections:
\begin{equation}
\hat \rho \equiv n_\alpha n_\beta T^{\alpha\beta}\,, \qquad
S_i \equiv - h_{i\alpha} n_\beta T^{\alpha\beta}  \qquad {\rm and} \qquad
S_{ij} \equiv h_{i\alpha} h_{j\beta} T^{\alpha\beta} \,.
\end{equation}
where $n_\alpha \equiv g_{\alpha \beta} u^\beta_{\rm rf}$ is the covariant 4-velocity of the comoving rest frame where $g_{\alpha\beta}$ is the metric tensor from eq.~\cref{eq:ds2} and $u_{\rm rf}^\beta = (\alpha^{-1},0,0,0)$ is the 4-velocity of the reference frame. 

%
\subsection{Evolution equations in spherical symmetry}
%

Adopting the shiftless gauge $\beta_i = 0$, the most general spherically symmetric metric becomes:
\begin{equation}
{\rm d}s^2 = -\alpha^2 {\rm d}t^2 + \hrr {\rm d}r^2 + \htt({\rm d}\theta^2 + \sin^2\theta {\rm d}\phi^2)\,,   \label{sphericmet}
\end{equation}
Shiftless gauge reduces ${\rm d}_t \rightarrow \partial_t$, because the Lie-derivative ${\cal L}_\beta$ now vanishes. Eventually we shall also adopt a constant lapse gauge $\alpha\equiv 1$, but in this section we keep an arbitrary lapse function for more generality. It turns out to be useful to make several redefinitions of the fundamental variables. First, we replace the covariant extrinsic curvature tensor $K_{ij}$ by the mixed tensor ${K^i}_j$. In the spherically symmetric case there are only two independent and non-vanishing components: $\KRr$ and $\KTt$, which coincide with the negatives of radial- and transversal expansion rates. For simplicity and for better numerical behavior we also replace the diagonal three-metric components by the local space and time dependent scale factors:
\begin{equation}
\htt \equiv (\rat)^2   \quad {\rm and} \quad 
\hrr \equiv \ar^2,
\label{eq:new_metric_variables}
\end{equation}
Indeed, in $h_{ij}$-variables, the equations for extrinsic curvatures contain several terms that each diverge at the origin of the coordinate system, but whose sum is finite and essential. Combining divergent terms leads to a significant loss of numerical accuracy and the crash of a code written in these variables. In the new local scale factor variables the problematic terms become summed analytically and the numerical issue vanishes. In addition we define the shorthand notations
\begin{equation}
X \equiv \frac{\ratr}{\ar } \quad {\rm and} \quad Z \equiv \frac{\alpha_{,r}}{\ar} \,. 
\end{equation}
With these definitions and variables the ADM equations for the metric~\cref{sphericmet} become:
\begin{align}
\dot a_r  &\equals - \alpha \ar \KRr 
\\
\dot a_\theta &\equals - \alpha \at \KTt
\label{eq:HrrEq}
\end{align}
and
\begin{align}
{\dot K}^r_{\;\;r} &\equals - \frac{Z_{,r}}{\ar} 
  + \alpha \left[ 2\KRr\KTt + (\KTt)^2 - \frac{2X_{,r}}{\rat\ar} \right]
  - \kappa {M^r}_r
\label{KRrEq} \\
{\dot K}^\theta_{\;\;\theta}  &\equals - \frac{XZ}{\rat}
  + \alpha \left[ \KRr\KTt + 2(\KTt)^2 - \frac{X_{,r}}{\rat\ar}
  + \frac{1 - X^2}{(\rat)^2} \right]- \kappa {M^\theta}_\theta \,.	
\label{KTtEq}
\end{align}
The only nontrivial momentum constraint~\cref{eq:C2} is the radial one with $i=r$:
\begin{equation}
\KTtr + \frac{\ratr}{\rat} (\KTt - \KRr) =  -\frac{\kappa}{2} S_r \,.
\label{momconst}
\end{equation}
Finally, the Hamiltonian constraint now reads:
\begin{equation}
 2 \KRr\KTt + (\KTt)^2 - \frac{2X_{,r}}{\rat\ar}
 + \frac{1 - X^2}{(\rat)^2}   =  \kappa \hat{\rho} \,. 
 \label{Hamconst}
\end{equation}
Either of the constraint equations~\cref{momconst,Hamconst} could be used to algebraically solve for $\KRr$ and substitute this in~\cref{eq:HrrEq,KTtEq}.  After $\KRr$ is thus eliminated, the other constraint equation remains as a useful check that can be used to monitor the accuracy of the numerical solution.

Equations~\cref{KRrEq,Hamconst} still contain numerically problematic terms, such as $(1-X^2)/(\rat)^2$, which appears to blow up at the origin, but in fact does not, because $X^2 \rightarrow 1$ for small $r$. However, adding the two divergent terms numerically yields a too large an error. To fix this, we decompose the radial scale function $\ar$ as in Ref.~\cite{Marra:2011zp} writing
\begin{equation}
   \ar \equiv \frac{\ratr}{\sqrt{1+2E}}.
\end{equation}
When we take $E$ as a new variable instead of $\ar$, we find that
\begin{equation}
   X = \sqrt{1+2E} \quad {\rm and} \quad \frac{1-X^2}{(\rat)^2} = -\frac{2E}{(\rat)^2} \,.
\end{equation}
That is, the numerically problematic terms combine to a single well behaved ratio (we will eventually absorb $r^2$ into $E$, which scales as $E \sim r^2$ for small radius). From Eq.~\cref{eq:HrrEq} we now find the evolution equation for $E$:
\begin{equation}
\dot{E} = \frac{\alpha}{2}\frac{\rat}{\ratr}(1+2E)\left( \frac{\kappa}{2}S_r -\frac{\alpha_{,r}}{\alpha} \KTt \right)\,,
\label{eq:Eequation}
\end{equation}
where we already used the momentum constraint to simplify the source term on the right hand side. 
                        
%
\subsection{Perfect multicomponent fluid}
%

Next we restrict our source into a set of coexisting spherically symmetric perfect-fluid components. The Energy-momentum tensor of a perfect multicomponent fluid is:
\begin{equation}
T^{\alpha\beta} = \sum_i \left( (\rho_i + p_i)u_i^\alpha u_i^\beta + p_i g^{\alpha\beta} \right) \,.
\label{eq:emomentum}
\end{equation}
where the pressure is given by some a priori known $t$ and $r$ dependent equation of state (EOS):
\begin{equation}
p_i \equiv w_i \rho_i \,.
\end{equation}
The four-velocity in our coordinate system is given by:
\begin{equation}
u_i^\alpha = (\frac{\gamma_{v_i}}{\alpha},\frac{\gamma_{v_i}v_i}{\ar},0,0)\,,
\end{equation}
where $v_i$ is the {\em proper} radial velocity of the $i$th fluid component and $\gamma_{v_i}^2 \equiv 1/(1-v_i^2)$ is the corresponding Lorentz gamma factor. 

%
\subsubsection{Fluid equations}
%

ADM equations for the metric still need to be complemented by evolution equations for the velocity $v_i$ and density fields $\rho_i$ of the individual fluids. These are obtained from the evolution equations for density and pressure components, which follow from the covariant conservation of the individual energy-momentum tensors $D_\mu T_i^{\mu\nu}= 0$.  One finds~\cite{Marra:2011zp}
\begin{align}
\frac{{\rm d}\rho_i}{{\rm d}\tau} &\equals -\Theta_i (\rho_i + p_i)
\nonumber \\
\frac{{\rm d}p_i}{{\rm d}\sigma} &\equals -a_i (\rho_i + p_i)\,.
\label{eq:rhopequations1}
\end{align}

\noindent Here the convective derivatives with respect to velocity and acceleration are defined as
\vskip-0.4truecm
\begin{align}
\frac{{\rm d}}{{\rm d}\tau} &\isequiv u_i^\alpha \partial_\alpha =
\frac{\gamma_{v_i}}{\alpha} \frac{\partial}{\partial t} 
            + \frac{\gamma_{v_i} v_i}{\ar} \frac{\partial}{\partial r} 
\nonumber \\
\frac{{\rm d}}{{\rm d}\sigma}  &\isequiv  a_i^\alpha \partial_\alpha =
\frac{\gamma_{v_i} v_i}{\alpha} \frac{\partial}{\partial t} 
            + \frac{\gamma_{v_i}}{\ar} \frac{\partial}{\partial r}
\label{eq:tauandsigmaderivatives}
\end{align}

\noindent and the expansion and the accelerator scalars $\Theta_i$ and $a_i$ are given by
\vskip-0.4truecm
\begin{align}
\Theta_i     \equals \Theta_{Ti} + \Theta_{Ri} + \Theta_{Ai}
\qquad {\rm and} \qquad
a_i          \equals  \frac{1}{v_i}\Theta_{Ti} + v_i\Theta_{Ri} \,,
\end{align}
\vskip-0.2truecm
\noindent where
\begin{align}
\Theta_{Ti} &\equals   \;\frac{v_i\gamma_{v_i}}{\alpha} \Big( \gamma_{v_i}^2 \dot v_i + \frac{\alpha'}{\ar}\Big)
\nonumber \\[-1ex]
\Theta_{Ri} &\equals \; \frac{\gamma_{v_i}^3 v_i'}{\ar} - \gamma_{v_i} \KRr
\nonumber \\[-1ex]
\Theta_{Ai} &\equals  -2\gamma_{v_i}\KTt + 2\gamma_{v_i}v_i\frac{X}{\rat} \,.
\end{align}
After some algebra~\cref{eq:rhopequations1} can be turned into equations for the fluid velocities and densities:
\begin{align}
   \dot v_i &\equals -\frac{\alpha'}{\gamma_{v_i}^2\ar} + \alpha v_i \frac{\gamma_{cv_i}^2}{\gamma_{v_i}^2} 
  \Big( \KRr - {\hat c}_{si}^2K
      +   \frac{2{\hat c}_{si}^2v_iX}{\rat} - \frac{\gamma^2_{v_i}v_i'}{\gamma_{c_i}^2\ar}\Big)
    - \frac{1}{1+w_i}\frac{B_i}{\gamma_{v_i}^2}, 
\label{eq:finalviequation} \\
   \frac{\dot \rho_{i}}{\rho_i} &\equals 
      - \alpha (1+w_i)\gamma_{cv_i}^2 \Big( - K + v_i^2 \KRr 
        + \frac{2v_iX}{\rat} + \frac{v_i'}{\ar} \Big)  
        - \frac{\alpha v_i}{\ar}\frac{\rho_i'}{\rho_i} + v_iB_i \,.
\label{eq:finalrhoiequation}
\end{align}
In these equations we denote
\begin{equation}
   {\hat c}_{si}^2 \equiv \Big(\frac{\partial p_i}{\partial \rho_i}\Big)_{t,r} \,,
\label{eq:sound_velocity}
\end{equation}
where both the time and radial coordinates were kept constant. This quantity arises from computing the derivatives~\cref{eq:tauandsigmaderivatives} using the chain rule. It coincides with the local sound speed squared in the case of a barotropic fluid, and we shall be referring to ${\hat c}_{si}$ as sound speed, even in cases where this is not the strictly correct identification. In addition we define the gamma-factors $\gamma_{c_i}^2 \equiv 1/(1-{\hat c}_{si}^2)$ and $\gamma_{cv_i}^2 \equiv 1/(1-v_i^2{\hat c}_{si}^2)$, as well as  $w_i \equiv p_i/\rho_i$ and 
\begin{equation}
B_i \isequiv \gamma_{cv_i}^2 
\left(v_i{\dot w}_{i\rho} +\frac{w_{i\rho}'}{\ar} + \frac{\alpha {\hat c}_{si}^2}{\gamma_{v_i}^2\ar} \frac{\rho_i'}{\rho_i}\right) \,,
\label{eq:Fiequation}
\end{equation}
where ${\dot w}_{i\rho}\equiv (\partial w_i/\partial t)_{\rho,r}$ and $w_{i\rho}' \equiv (\partial w_i/\partial r)_{\rho,t}$ are partial time and radial derivatives computed at a constant $\rho_i$. In the examples considered in this text $w_{i\rho}'$ always vanishes. Note that for $\alpha'=0$, the only terms	not proportional to $v_i$ in the velocity equation are the second and third terms in the $B_i$-function, which on the other hand are proportional both to sound speed and density gradient or the explicit radial dependence of the equation of state. If these terms vanish, the velocity remains zero at all times as it should. Then also all fluid corrections to the density equation disappear and $\rho_i$ becomes solvable analytically as a function of metric components (see eq.~\cref{eq:dustdensity} below).

%
\subsubsection{Source Terms}
%

Finally, we need to work out explicit expressions for the radially symmetric source terms in Eqs.~\cref{eq:HrrEq,Hamconst} and~\cref{eq:Eequation}. In the frame we have chosen $n_\alpha = g_{\alpha \beta} u^\beta_{\rm rf} = (-\alpha;{\bf 0})$. Thus for any diagonal system with a vanishing shift function we get:
\begin{equation}
\hat \rho =  \alpha^2T^{tt} =  -{T^t}_t\,, \qquad
S_i = \alpha {T^t}_i  \qquad {\rm and} \qquad
{S^i}_j = {T^i}_j
\end{equation}
from which follows
\begin{equation}
{M^i}_i = \frac{1}{2} \Big( {T^i}_i - {T^t}_t - \sum_{k\neq i}{T^k}_k \Big)\,.
\end{equation}
Because ${T^\theta}_\theta = {T^\phi}_\phi$ we find in particular that:
\begin{equation}
{M^r}_r = \frac{1}{2} \Big( {{T^r}_r - T^t}_t - 2 {T^\theta}_\theta \Big)
\quad {\rm and} \quad
{M^\theta}_\theta = -\frac{1}{2} \Big( {T^r}_r + {T^t}_t \Big)\,.
\end{equation}
In our coordinate system we can easily work out the nontrivial components ${T^t}_t = \sum$, ${T^r}_r$ and ${T^\theta}_\theta$. They are listed in appendix~\cref{appendix:B}. From them we obtain the necessary source terms:
\begin{align}
\hat \rho &\equals  \sum_i \gamma_{v_i}^2(\rho_i+v_i^2p_i)
\nonumber\\[-1.1ex]
S_r &\equals \ar \sum_i (\rho_i+p_i)v_i\gamma_{v_i}^2 \,.
\nonumber\\[-1.3ex]
{M^r}_r &\equals \frac{1}{2} \sum_i \Big[ \rho_i - p_i + 2(\rho_i + p_i)v_i^2\gamma_{v_i}^2 \big]
\nonumber\\[-1.3ex]
{M^\theta}_\theta &\equals \frac{1}{2} \sum_i (\rho_i - p_i)\,.
\end{align}
This completes our task of writing explicitly a closed set of equations for the metric components, extrinsic curvatures and fluid velocities and densities in a general shiftless gauge. Next we need to set up physically sensible initial- and boundary conditions.

%
\subsection{Boundary conditions}
\label{sec:bndrycond}
%

We choose the boundary conditions such that each variable agrees with the corresponding background solution at the ``grid boundary" $r=r_{gb}$. We will consider only the flat FLRW background in this paper, so that the corresponding boundary conditions for the metric and extrinsic curvature variables are:
\begin{equation}
\ar(t,r_{gb}) = \at(t,r_{gb}) = \bar a(t), \quad {\rm and} \quad
{K}^r_{\,\,\,r}(t,r_{gb}) = {K}^{\theta}_{\,\,\,\theta}(t,r_{gb}) = - \bar{H}\,,
\label{eq:boundary1}
\end{equation}
where $\bar a$ and $\bar H$ are the background FLRW scale factor and Hubble rate. Note that we will frequently use the background $\bar a$ as the time variable of our computations. The above conditions imply that $E(t,r_{gb})=0$. Similarly for densities and velocities:
\begin{equation}
\rho_i(t,r_{gb}) = \bar \rho_i(t) \quad {\rm and} \quad v_i(t,r_{gb}) = 0 \,.
\label{eq:boundary2}
\end{equation}

We note that this embedding of an inhomogeneity into the FLRW metric can cause numerical issues. The problem shows up when the acoustic sound waves, induced by pressure gradients, are reflected back from $r=r_{gb}$ by our boundary conditions, and reach the origin where they cause numerical stability problems. Indeed, conditions~\cref{eq:boundary1,eq:boundary2} constitute a case in which our bubble touches another, identical spherical bubble at every point on the surface $r=r_{gb}$. The reflected waves are then equivalent to a coherent sum of incoming sound waves waves from infinitely many surrounding bubbles. This unphysical situation can be ameliorated in several ways. First, one can make sure that $r_{gb}$ is so much larger than the actual size of the bubble, so that reflected waves never make it back before the end of the calculation. Another possibility is to use $r_{gb}$ only somewhat larger than $r_b$, and judiciously induce diffusion in region $r_b \ll r < r_{gb}$ that damps the sound waves without changing the solution in the region of interest. We have studied both approaches with well agreeing results.

%
\subsection{Initial conditions}
\label{sec:initial_conditions}
%

Setting up consistent initial conditions requires a little more work and definitions.
First, because of the spherical symmetry and freedom of choosing the units, we can always set the angular scale factor such that~\cite{Marra:2011zp}:
\begin{equation}
\at^{\rm in} \equiv \at(t_{\rm in},r) \equiv \bar a(t_\text{in}) \equiv \bar a_{\rm in}\,.
\label{eq:InCforHtt}
\end{equation}
where $t_\text{in}$ is the starting time of the computation. For all other variables, it is important to give initial conditions at a time $t_\text{in}$ when inhomogeneity is still well outside the horizon. For now, we will also assume that the perturbation is adiabatic. Then, in the superhorizon scales, all metric and density variables eventually depend on just one arbitrary function, which we will choose to be the effective gravitating mass $F_M^{\rm in}$ of the dust component. To be specific, we define
\begin{equation}
\delta^{\rm in}_{F_M} \equiv \frac{F^{\rm in}_M}{F^{\rm in}_{M,\rm{bkg}}} - 1 = \bar{\delta}_M^{\rm in} g_F(r). 
\label{FbarEq}
\end{equation}
where $\bar \delta_M$ is the amplitude of the initial gravitating mass perturbation, the background gravitating mass is $F^{\rm in}_{M,\rm{bkg}} = (4\pi/3) (\rat^{\rm in})^3 \bar{\rho}_i^{\rm in} =(4\pi/3) (\bar a_{\rm in} r)^3 \bar{\rho}_i^{\rm in}$, and $g_F(r)$ is an arbitrary initial density profile. In all examples of this paper we will use the profile\footnote{With our default parameters~\cref{densprof} is identical to the profile used in~\cite{Marra:2011zp}, up to a few per cent difference in the overall normalization, which we have taken into account in our explicit comparisons.}
\begin{equation}
    g_F(r) \equiv \frac{1 - \tanh((r - r_b)/\Delta r)}{1 + \tanh(r_b/\Delta r)}
    \label{densprof}
\end{equation}
The radius $r_b$ determines the size of the inhomogeneity and $\Delta r$ controls the steepness of transition to homogeneity. In all applications below, we will use the value $\Delta r=0.15\,r_b$. For $r-r_b \gg \Delta r$ the perturbation vanishes and the densities reduce to their FLRW values. Please note that we always parametrize the size of the inhomogeneity by its {\em radius} $r_b$, whereas one usually refers the void size by its diameter $d$. In this sense, for us {\em e.g.}~a $100$ Mpc void is a bubble with a radius $r_b = 50$ Mpc.

As long as the perturbation is well outside horizon, the gravitating mass of the dust $F_M$ and the dust density $\rho_M$ are related by $\partial_r F_M = 4\pi r^2 \rho_M$. Then differentiating~\cref{FbarEq} with respect to $r$, we immediately obtain for the density perturbation in the dust:
\begin{equation}
\delta^{\rm in}_{ \rho_M} \; \equiv \; \frac{\rho_M^{\rm in}}{\bar \rho_M^{\rm in}}-1 
\; = \; \bar \delta_M^{\rm in}\Big( 1 + \frac{r}{3}\partial_r\Big)  g_F(r). 
\label{rhobarEq}
\end{equation}
For adiabatic superhorizon scale perturbations, the amplitude of all other fluid perturbations are related to that of the dust component by~\cite{Ballesteros:2010ks}:
\begin{equation}
{\bar\delta}_i^{\rm in} = \bar \delta_M^{\rm in} (1+\bar w_i) \frac{5-6\bar c_{ai}^2}{5-15\bar w_i + 9\bar c_{ai}^2}\,,
\label{eq:ballesteros}
\end{equation}
where and $\bar w_i$ and $\bar c_{ai}$ are the background EOS~parameter and the adiabatic sound speed of the $i$th fluid component. We assume that $\bar w_i$'s are constant for $t<t_{\rm in}$, from which it follows that $\bar c_{ai} = {\hat c}_{si}$. However, for the sake of completeness we will allow independent profiles for all initial fluid perturbations, and the above discussion generalizes to 
\begin{equation}
\delta^{\rm in}_{ \rho_i} = \bar \delta_i^{\rm in} \, \Big( 1 + \frac{r}{3}\partial_r\Big)g_{F,i}(r) \hspace*{10pt} \mathrm{and} \hspace*{10pt}   \delta^{\mathrm{in}}_{F_i}(r) =  \bar{\delta}^{\mathrm{in}}_i g_{F,i}(r).
\label{adcond}
\end{equation}
Since the peculiar motions should be negligible at large scales at very early times, it is usually sufficient to adopt vanishing initial velocities for all fluids:
\begin{equation}
v_i^{\rm in}\equiv 0 \,.
\end{equation}
We adopt this choice in the examples of next chapter. A more refined initial condition could also be implemented following {\em e.g.}~\cite{Batista_2023}.

\paragraph{Simultaneous bang-time and a vanishing lapse function.}

Next, we need to set initial conditions for the remaining metric components consistent with the matter perturbations and our choice for $\at^{\rm in}$. Eq.~\cref{eq:InCforHtt} already sets $\ar^{\rm in} = \bar a_{\rm in}/\sqrt{1+2 E_{\rm in}}$ and $\KRr$ can be solved from one of the constraint equations at all times, so we only need the initial conditions for $\KTt$ and $E$. Both of these functions are connected to the total gravitational mass $F$ through the angular expansion rate equation~\cite{Marra:2011zp}:
\begin{equation}
(\KTt)^2 = \frac{1}{\alpha^2}\Big(\frac{{\dot a}_\theta}{\at}\Big)^2 
=\frac{2G_NF}{(\rat)^3} +  \frac{2E}{(\rat)^2}\,.
\label{eq:bcondforktt}
\end{equation}
We wish to set the initial conditions to correspond to simultaneous big-bang singularity in the coordinate time $t=0$. This ensures that there are no decaying modes. This makes sense if initial conditions are given well after inflation, as any decay modes persisting long after inflation would have to be enormous initially, and hence in conflict with the inflationary paradigm \cite{Zibin:2008vj}. The equation for vanishing bang-time can be found by formally integrating Eq.~\cref{eq:bcondforktt}:
\begin{equation}
t_{\rm in} 
\equiv \int^{t_{\rm in}}_0  \frac{ \dot a_\theta{\rm d}t}{\alpha \at\KTt}      
\, = \;  \int^{t_{\rm in}}_0  \frac{ \dot a_\theta{\rm d}t}{\alpha \at \sqrt{\frac{2G_N F}{ (\rat)^3}+\frac{2E}{(\rat)^2}}}.      
\label{tbarint}
\end{equation}
The problem here is that, unlike in the pure LTB model, the lapse $\alpha$, the effective gravitational mass $F$ and the curvature function $E$ all depend on time, and hence equation~\cref{tbarint} cannot be integrated without further assumptions. Therefore, we first go to constant lapse gauge $\alpha\equiv 1$. Second, we assume that equation of state parameters $w_i$ are roughly constant and equal to their background FLRW values for $t<t_{\rm in}$, and we specify initial conditions at sufficiently early so that the pressure gradients can be neglected. Then the evolution of independent causal patches effectively followed an FLRW model, approximately setting $E(t<t_{\rm in},r)\approx E_{\rm in}$. This observation gives scaling relations for the individual fluid components, so that we can approximate the time dependence of $F$ as\footnote{Equation~\cref{Fbarsum} follows from the fact that at superhorizon scales $F'\sim \rho$ and from solving Eq.~\cref{eq:finalrhoiequation} in the limit $v_i=0$ and $\alpha = 1$ and using $\at \sim \ar \sim \bar a$.}
\begin{equation}
F \approx  \sum_{i=1}^n \left(\frac{\at^{\rm in}}{\at }\right)^{3w_i} F^{\rm in}_i,      
\label{Fbarsum}
\end{equation}  
where $F^{\rm in}_i$ is the initial profile for the $i$th component gravitating mass. While there is no really meaningful division into independent components for the gravitating mass in general, such a division can be made at a given surface, which here is chosen to be the initial one. Indeed, if we trace back our argument leading from~\cref{FbarEq} to~\cref{rhobarEq} backwards for arbitrary fluid component $i$, we find first
\begin{equation}
2G_N F^{\rm in}_i = 2G_NF^{\rm in}_{\rm{bkg},i}( 1 + {\delta}_{F_i}^{\rm in} ) = \bar H_{\rm in}^2\bar \Omega_i^{\rm in}( 1 + {\delta}_{F_i}^{\rm in} )(\rat^{\rm in})^3 \,,
\label{eq:Fincond}
\end{equation}
where $\bar H$ is the background expansion rate, $\bar \Omega_i \equiv \bar \rho_i/\bar\rho_{\rm cr}$ and, $\delta_{F_i}^{\rm in} \equiv \bar\delta_i^{\rm in}g_F(r)$ in accordance  with~\cref{FbarEq}. Then from~\cref{eq:bcondforktt} and~\cref{Fbarsum} we find that at all times $t<t_{\rm in}$:
\begin{equation}
\KTt \approx \bar H_{\rm in}\, \frac{\at^{\rm in}}{\at}\,
\left[ \sum_{i=1}^n \bar\Omega_i^{\rm in}(1 + \delta_{F_i}^{\rm in})
\Big(\frac{\at^{\rm in}}{\at}\Big)^{3w_i+1} + \frac{2E_{\rm in}}{(\bar a\bar H)_{\rm in}^2} \right]^{1/2},
\label{FbarEq2}
\end{equation}
where all time dependence comes through dependence in $\at$. Here $\bar a$ and $\bar H$ are the scale factor and the expansion rate of the homogeneous background FLRW spacetime. We will also frequently parametrize functions of time in terms of a background redshift $\bar z \equiv \bar a_0/\bar a - 1$. 

Making a change of variables $t \rightarrow y \equiv \at/\at^{\rm in}$ and defining $E_{\rm in} \equiv (\bar a \bar H r)_{\rm in}^2 \epsilon_{\rm in}$, we can write equation~\cref{tbarint} as
\begin{equation}
\bar H_{\rm in} t_{\rm in} \approx \int_0^{1}\frac{{\rm d}y}{\sqrt{\sum_{i=1}^n 
\bar\Omega_i^{\rm in}(1 + \delta_{F_i}^{\rm in})y^{-3w_i-1} + 2\epsilon_{\rm in}}}.   
\label{tbarEq}
\end{equation}
This equation holds true for all $\delta_{F_i}$ and $\epsilon_{\rm in}$, and in particular in the homogeneous background where $\delta_{F_i} = \epsilon_{\rm in} = 0$, where it reduces to the standard time integral in the FLRW space. Thus, subtracting the integral\footnote{Note that with some choices of $w_i$, such as if the system contains only dark energy with no matter component, integral~\cref{tbarEq} can diverge. In such cases the initial conditions need to be considered more carefully depending on the specific setup.} for the homogeneous case from the inhomogeneous one leads to a relation between the gravitating mass inhomogeneities $\smash{\delta^{\rm in}_{F_i}}$ and the metric perturbation $e_{\rm in}$. Expanding the integrand in the expression~\cref{tbarEq} to first order in small quantities $\epsilon_{\rm in}$ and $\smash{\delta_{F_i}^{\rm in}}$ then gives with:
\begin{equation}
\epsilon_{\rm in}
= \frac{E_{\rm in}}{(\bar a\bar Hr)_{\rm in}^2} 
\,\approx \,
 - \frac{1}{2}\sum_i \left[\bar\Omega_i^{\rm in} \bar\delta_i^{\rm in} R_i^{\rm in} g_{F,i}(r)\right] \,  
\label{Eapp}
\end{equation}
where
\begin{equation}
R_i \equiv \frac{I_{3w_i+1}}{I_{0}}, \qquad {\rm with} \qquad   
I_p \equiv \int_0^1 \frac{{\rm d}yy^{-p}}{\left(\sum_{i=1}^n{\bar\Omega}^{\rm in}_iy^{-3w_i-1}\right)^{3/2}} \,.
\label{eq:RandIvariables}
\end{equation}
This result is a generalization of the result of~\cite{VanAcoleyen:2008cy} for the LTB model. Indeed it is easy to see that in the pure dust case Eq.~\cref{Eapp} reduces to the result
\begin{equation}
\epsilon_{\rm in}=-\frac{5}{6} \bar\delta^{\rm in}_M \,g_F(r) \,,
\label{Edust}
\end{equation}
while in the pure radiation case one finds instead
\begin{equation}
\epsilon_{\rm in} = -\bar\delta^{\rm in}_{\rm{rad}}\, g_F(r).
\end{equation}
Finally, with the solution~\cref{Eapp} at hand, we get $(\KTt)_{\rm in}$ from~\cref{FbarEq3}. Again working to the first order accuracy in perturbations we find:
\begin{equation}
\frac{(\KTt)_{\rm in}}{\bar H_{\rm in}} \, \approx \, 1+ \frac{1}{2}\sum_{i=1}^n \left[\bar\Omega_i^{\rm in} \bar\delta_{i}^{\rm in}(1 -  R_i^{\rm in})  g_{F,i}(r) \right]\,  
\,.
\label{FbarEq3}
\end{equation}
As promised, all fluid densities and metric components have now been expressed in terms of the single set of functions $g_{F,i}(r)$, determining the inhomogeneity at the appropriately chosen initial time $t_{\rm in}$.  To connect the initial values to present day observables, we note that $\bar \Omega_{i}$'s at arbitrary background redshift $\bar z$ are related to the current values $\Omega_{i0}$ through equation 
\begin{equation}
\bar \Omega_{i} = \frac{\Omega_{i0}}{\sum_{k=1}^n\Omega_{k0}(1+\bar z)^{3(\bar w_k-\bar w_i)}}\,.
\end{equation}
A generalization for time-dependent EOS parameters is shown in~\cref{eq:OmegaBar_timedep}. Finally, let us remark that in $\alpha = 1$ gauge the fluid equations for the dust can be solved analytically.  Indeed, in this gauge the rest-frame coincides with the (pressureless) comoving dust frame, so that the dust velocity vanishes identically:
\begin{equation}
v_{M}(t,r) \equiv 0.
\label{eq:dustVelo}
\end{equation}
Moreover, given zero velocity, the equation~\cref{eq:finalrhoiequation} for the dust density reduces to:
\begin{equation}
\frac{\dot \rho_M}{\rho_M} = \frac{\dar}{\ar} + 2 \frac{\dat}{\at} \,.
\label{eq:dustdens}
\end{equation}
This can be immediately integrated to give 
\begin{equation}
\rho_M(t,r) = 
\frac{\ar^{\rm in}}{\ar}\Big(\frac{\at^{\rm in}}{\at}\Big)^2 \rho_M^{\rm in}  \,.
\label{eq:dustdensity}
\end{equation}
That is, we obtain $\rho_M$ from its initial condition and the known solutions to the metric components.

%
\subsection{Light propagation}
%

The geodesic equation for photons can be written as $k^{\mu}\nabla_{\mu}k^{\nu} = 0$, where the wave vector $k^{\mu}$, tangent to the path $x^{\mu}\left(\lambda\right)$, satisfies the null condition $k_{\mu}k^{\mu} = 0$. Here $\lambda$ is an affine parameter labeling the points along the photon geodesic. Because the direction of the angular momentum is conserved, one can choose the coordinate $\theta = \frac{\pi}{2}$, thus
\begin{equation}
k^{\mu} = \left( \frac{\mathrm{d} x^{t}}{\rm{d} \lambda}, \frac{\mathrm{d} x^{r}}{\mathrm{d}\lambda}, 0, \frac{\mathrm{d} x^{\phi}}{\rm{d} \lambda} \right).
\end{equation}
In practice, we can solve for $(k^0)^2 = \alpha^{-2} h_{ij} k^i k^j$ from the null condition and write the geodesic equations for the spatial components as
\begin{equation}
    \frac{dk_i}{dt} = -\frac12 k^0 \partial_i \alpha^2 + \frac12 (k^0)^{-1} k^n k^m \partial_i h_{mn},
\end{equation}
where we have again set $\beta^i = 0$. Additionally, the conservation of the magnitude of the angular momentum allows one to set $k_{\phi} =  k^{\phi}r^2\at^2\sin^2{\theta} = \; $const. These equations can then be integrated alongside the definition of the momentum vector $dx^i/dt = (k^0)^{-1} k^i$. The true redshift (in contrast to the background redshift $\bar z$) for comoving observers and sources is then obtained simply by comparing the frequencies:
\begin{equation}
    z = \frac{k^0_{\text{src}}}{k^0_{\text{obs}}} - 1.
\end{equation}

A complete description of light propagation in an inhomogeneous universe involves also the evolution of the cross section of a photon beam. To this end we first need to construct a coordinate system along the geodesic. Two four-vectors $s^{\mu}_{\alpha}$, where $\alpha = \{1,2\}$, can be defined to be perpendicular to the geodesic $g_{\mu\nu}k^{\mu}s^{\nu}_{\alpha} = 0$. Additionally, one more basis vector $l^{\mu}$ can be introduced, for which the only requirement is $k_{\mu}l^{\mu} = -1$. Eventually, a complete basis set of 4-vectors that fulfill the above conditions can be chosen as
\begin{equation}
\begin{split}
s^{\mu}_1 &= \left( 0, 0, \frac{1}{\rat }, 0\right), \hspace*{140pt} k^{\mu} = \left( k^t, k^r, 0 \,, \,k^{\phi} \right),\\
s^{\mu}_2 &= \left( f_t\left(k^{\mu}, g_{\mu\nu}\right), \, f_r\left(k^{\mu}, g_{\mu\nu}\right), \,0, \,f_{\phi}\left(k^{\mu}, g_{\mu\nu}\right)\right), \hspace*{14pt}l^{\mu} = \left( l^t, l^r, l^{\theta}, l^{\phi} \right),
\label{eq:basis}
\end{split}
\end{equation}
where $f_t, f_r$ and $f_{\phi}$ are yet arbitrary functions of the wave vector and the metric tensor.

When a beam of light passes through an inhomogeneous density distribution, it experiences expansion and shear. As long as the curvature radius of the spacetime is much larger than the wavelength of the photons, this phenomenon can be mathematically described with the Sachs optical equations \cite{Sachs:1961zz}
\begin{align}
    \frac{d\theta}{d\lambda} &= -\frac12 \theta^2 - 2\sigma^2 - R_{\mu\nu} k^\mu k^\nu, \label{eq:sachs1}\\
    \frac{d\sigma_{AB}}{d\lambda} &= -\theta\sigma_{AB} - \left[R_{\alpha\mu\beta\nu} s_A^\alpha s_B^\beta - \frac12 \delta_{AB} R_{\mu\nu}  \right] k^\mu k^\nu, \label{eq:sachs2} 
\end{align}
where $\theta$ is the expansion scalar and $\sigma_{AB}$ the shear tensor
\begin{equation}
\sigma_{AB} =
  \begin{bmatrix}
     \sigma_{11}       & \quad\sigma_{12} \\
     \sigma_{12}       & \quad-\sigma_{11} \\
  \end{bmatrix},
\label{eq:shear}
\end{equation}
where $\sigma^2 = \sigma_{11}^2 + \sigma_{12}^2$. The angular diameter distance can then be obtained from the expansion scalar as
\begin{equation}
    \frac{d \text{ln} D_A }{d\lambda} = \frac{1}{2} \theta. \label{eq:angular_distance}
\end{equation}
The frame vectors $s_A^\mu$ should be evolved alongside this system according to their parallel transport equations $k^\mu \nabla_\mu s_A^\alpha = 0$. The various tensors necessary for a practical implementation of equations~\cref{eq:sachs1,eq:angular_distance} can be found in the appendix.

The initial conditions for the optical equations can be determined by demanding that the local evolution at the ray origin point matches the evolution in a FRW-spacetime with a matching local expansion rate. The expansion scalar diverges at $D_A = 0$, which poses a problem for setting the initial conditions for $D_A$ and $\theta$. We sidestep this issue by setting $\theta$ and $D_A$ to their FRW-values
\begin{equation}
    \theta(z) = 2 k^0 \left( \frac{\dot a(z)}{a(z)} - \frac{1}{D_A(z)} \right)
    \quad {\rm and} \quad
    D_A(z) = a(z) \int_{a_{obs}}^{a(z)} \frac{da}{a \dot a(z)}
    \label{eq:DA_semi_analytic}
\end{equation}
for the first few timesteps, where $a(z)$ is calculated as in the background FRW-universe. The error induced by this approximation will be discussed in appendix~\cref{appendix:sachs_equations}. We set the initial shear matrix and $s_2^0$ to zero and solve for the remaining components of the initial $s_2^\mu$ by requiring it to be orthogonal to the other three vectors defined in~\cref{eq:basis}. 

%
\section{Examples}
%

We shall now demonstrate the code by solving our equations in several, successively more complicated systems. To test the numerical setup, we first consider the simple pure dust LTB-model embedded in a flat EDS-background. Second, we consider models including dust and another fluid component with a generic equation of state, which includes matter and radiation and different exotic dark energy models with spatially varying sound speed as limiting cases. In particular we solve the case of an exotic DE-fluid with a non-negligible sound speed, a solution which has previously suffered from numerical problems~\cite{Marra:2011zp}. Third, we consider a model with dust and radiation and discuss issues arising with fluids with very large sound speeds. Fourth, we demonstrate the computation of distance observables by integrating over light rays on a metric corresponding to a model with CDM, DE and radiation fluids. Finally, we extend the base equations and code to systems with interacting fluids. In particular we study a case with interacting dark matter and dark energy fluids.  We will be always working in the synchronous gauge $\alpha \equiv 1$, which we already chose in specifying the initial conditions for metric components. In addition we use the momentum constraint to solve for $\KRr$, and finally perform several redefinitions and re-scalings of both dependent and independent variables to move into simple dimensionless variables and coordinates. We give the explicit equations of motion and initial and boundary conditions in these variables in appendix A.

%
\subsection{Comparison  to the standard LTB-model}
%

\begin{figure}
    \begin{center}
    \includegraphics[width=0.9\linewidth]{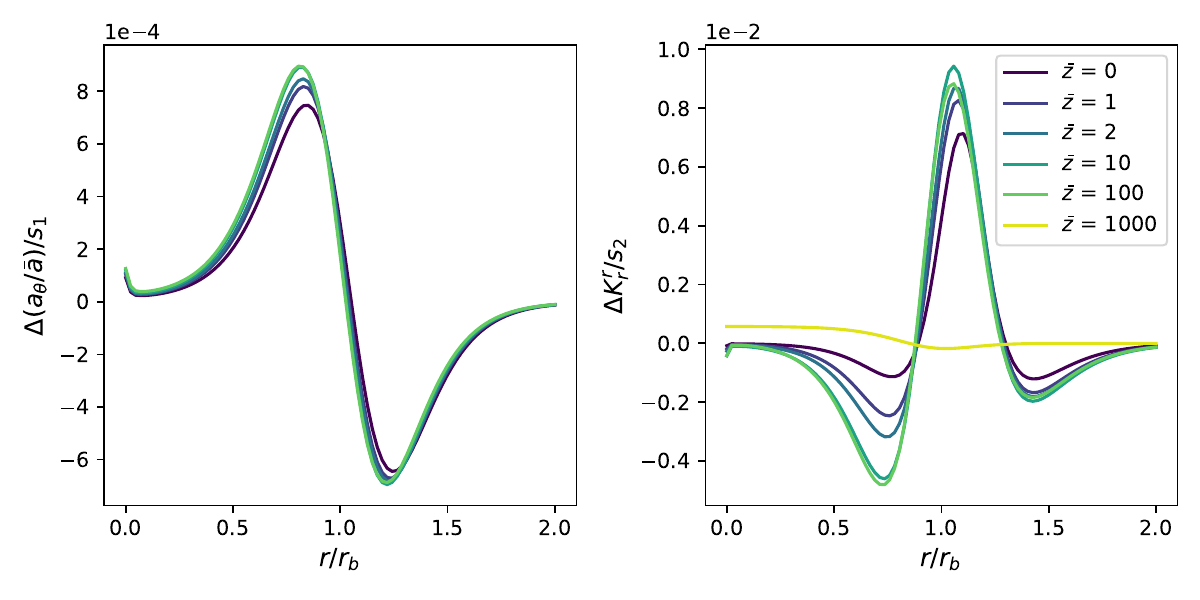}
    \caption{Left panel: The difference in the scaled metric component $\at$ between numerical and analytical solutions in a pure dust setup, scaled with the central value of the metric component. Right panel: The difference between the extrinsic curvatures $(K^r_{\phantom{a}r})^{\text{MC}}$ and $(K^r_{\phantom{a}r})^{\text{HC}}$ solved from the momentum and Hamiltonian constraints~\cref{momconst,Hamconst} in the same setup. The differences are scaled with the maximum perturbations $s_1 = \text{max}(a_\theta/\bar a - 1)$ and $s_2 = \text{max}(K^r_{\phantom{a}r} + \bar H)$ on each time-slice.}
    \label{LTB1}
    \end{center}
\end{figure}

First we make a direct comparison between the standard LTB model and the corresponding solution produced with our code. In this case, only dust is present and the inhomogeneity is embedded in a flat Einstein-de Sitter background model. The exact solutions to Einstein equations are known and can be given in simple parametric forms. In the hyperbolic case ($E(r)>0)$ one finds:
\begin{equation}
\rat = \frac{G_NF}{2E}\Big( {\rm cosh}(u\sqrt{2E})-1\Big) \,,
\label{eq:Httexact}
\end{equation}
where  $E$ and $F$ are constants defined by Eqs.~\cref{Edust,eq:Fincond} with $\bar\Omega_{\DM}=1$, and
\begin{equation}
t = \frac{G_NF}{2E}\Big( \frac{1}{\sqrt{2E}}{\rm sinh}(u\sqrt{2E})-u\Big) \,.
\label{eq:texact}
\end{equation}
It is simple to solve $\at(t,r)$ from these equations and compare the result with a  numerical solution. Here we are only interested in the accuracy of the numerical solution and not the solution itself.  So, in the left panel of figure~\cref{LTB1} we plot the deviation between the numerical and exact solutions for the angular scale function $\at$, normalized to exact $\at(r\!=\!0)$. For the inhomogeneity we used $r_b = 50$ Mpc and $\bar{\delta}_{\DM}^{\rm in}=-1.5\times 10^{-3}$ at a background redshift $\bar z_{\rm in}=1000$. The numerical solution is seen to be accurate to one part in $10^3$, just as we would expect from using linearized initial conditions. The relative error shows clear numerical noise for the first few timesteps, but afterwards it settles down to a near constant profile. The errors are entirely controlled by numerical tolerances and could be reduced by increasing the accuracy of the solver or the initial conditions, or by setting the initial conditions at an earlier time. In particular, note that the relative error does not grow noticeably over the course of the run.

%
%

In the right panel of figure~\cref{LTB1} we plot the relative difference between the $\KRr$-function obtained from the momentum constraint, which was used to compute $\KRr$ in the code, and the same function obtained from the Hamiltonian constraint (for more details see appendix A). This difference is sensitive to the numerical accuracy of the solution independently of whether the exact solution is known. While the error initially grows as a function of time, we see that even at $z=0$ the Hamiltonian constraint is satisfied up to $10^{-2}$ precision in normalized units. Moreover, the relative error actually starts to decrease towards the end of the run. We note that the per cent level errors in Hamiltonian constraint at the end of the run correspond to sub-per-mille level errors in metric components. We monitored the numerical accuracy of our results by keeping track of the Hamiltonian constraint in this manner in all simulations in this paper. In all cases errors remained small and the constraint behaved similarly to the standard LTB-case shown in Fig.~\cref{LTB1}.

%
\subsection{Dark Energy with a varying sound speed}
\label{sec:de_varying_sound_speed}
%

As our second example, we consider the case of two ideal fluids, one of which is dust denoted by $\rho_{\DM}$, with an equation of state $w_{\DM} = 0$, and the other is a generic fluid with an EOS\footnote{In Eq.~\cref{weq}, we have encoded the time dependence of the $w$ parameter through a time-dependent generic fluid. However, we may directly encode this time dependence in $w$. This is particularly relevant considering new observational datasets, such as those provided by the Dark Energy Spectroscopic Instrument Data Release 2~\cite{DESI:2025zgx}. DESI DR2 has enabled tight constraints on the expansion history of the universe and the growth of cosmic structure, providing background and perturbation observables critical for distinguishing between constant and evolving dark energy models. The observational results from DESI DR2 also show that simple constant-$w$ models are increasingly disfavored in favor of those allowing a transition in the equation of state~\cite{Shah:2024rme}. In particular, some analyses~\cite{Zhao:2017cud, Zhai:2023yny} report a mild preference for $w(\bar z) < -1$ at redshifts $\bar z\geq 0.5$ and $w(\bar z) >-1$ at lower redshifts. This indicates a crossing of the so-called phantom divide line $w(\bar z) = -1$, a phenomenon that is complex to accommodate within standard single-field quintessence models~\cite{PhysRevD.71.023515, PhysRevD.71.047301}.}
\begin{equation}
w(\rho_{\DE},t,r) \equiv \bar w_{\DE}\left(\frac{\rho_{\DE}(t,r)}{\bar \rho_{\DE}(t)}\right)^{\beta_{\DE}-1}  \equiv w_{\DE}.
\label{weq}
\end{equation}
For $\beta_{\DE} = 1$, this EOS reduces to familiar cases: another dust component with $\bar w_{\DE} = 0$ and a radiation fluid for $\bar w_{\DE} = 1/3$. However, for $\beta_{\DE} \neq 1$,~\cref{weq} allows for a more general, density-dependent polytropic EOSs which, for $\bar w_{\DE}<-1/3$ describe generalized, density dependent DE-fluids. Eq.~\cref{weq} gives the squared sound speed
\begin{equation}
{\hat c}_{s,\DE}^2 = \beta_{\DE} w_{\DE} \,. 
\label{eq:csa}
\end{equation}
We assume that $\beta_{\DE} \le 0$ so that ${\hat c}_{s,\DE}^2$ is positive. Note that setting $\beta_{\DE}=0$ describes a system with a zero sound speed and a homogeneous pressure $p_{\DE}(t,r)=\bar p_{\DE}(t)$, even when the energy-density of the fluid is inhomogeneous. The constant-$\rho$ time-derivative of the EOS parameter appearing in function $B_i$ in fluid equations here becomes
\begin{equation}
{\dot w}_{{\DE}\rho} = -3(1-\beta_{\DE})(1+\bar w_{\DE}) \bar Hw_{\DE} \,,
\label{eq:wdot}
\end{equation}
where $\bar H$ is the background expansion rate of the universe. This explicit time dependence came entirely from the background energy density $\bar\rho(t)$. For the superhorizon scale bubble, the scale of the initial adiabatic DE-fluid perturbation is related to that of the DM-fluid by Eq.~\cref{eq:ballesteros}, which can be written as:
\begin{equation}
\bar \delta_{\DE}^{\rm in} = \bar \delta_{\DM} (1+\bar w_{\DE}) \frac{5-6\beta_{\DE} \bar w_{\DE}}{5-15\bar w_{\DE} + 9\beta_{\DE} \bar w_{\DE}}\,.   
\label{adcond}
\end{equation}
%
%
\begin{figure}
\begin{center}
    \includegraphics[width=\linewidth]{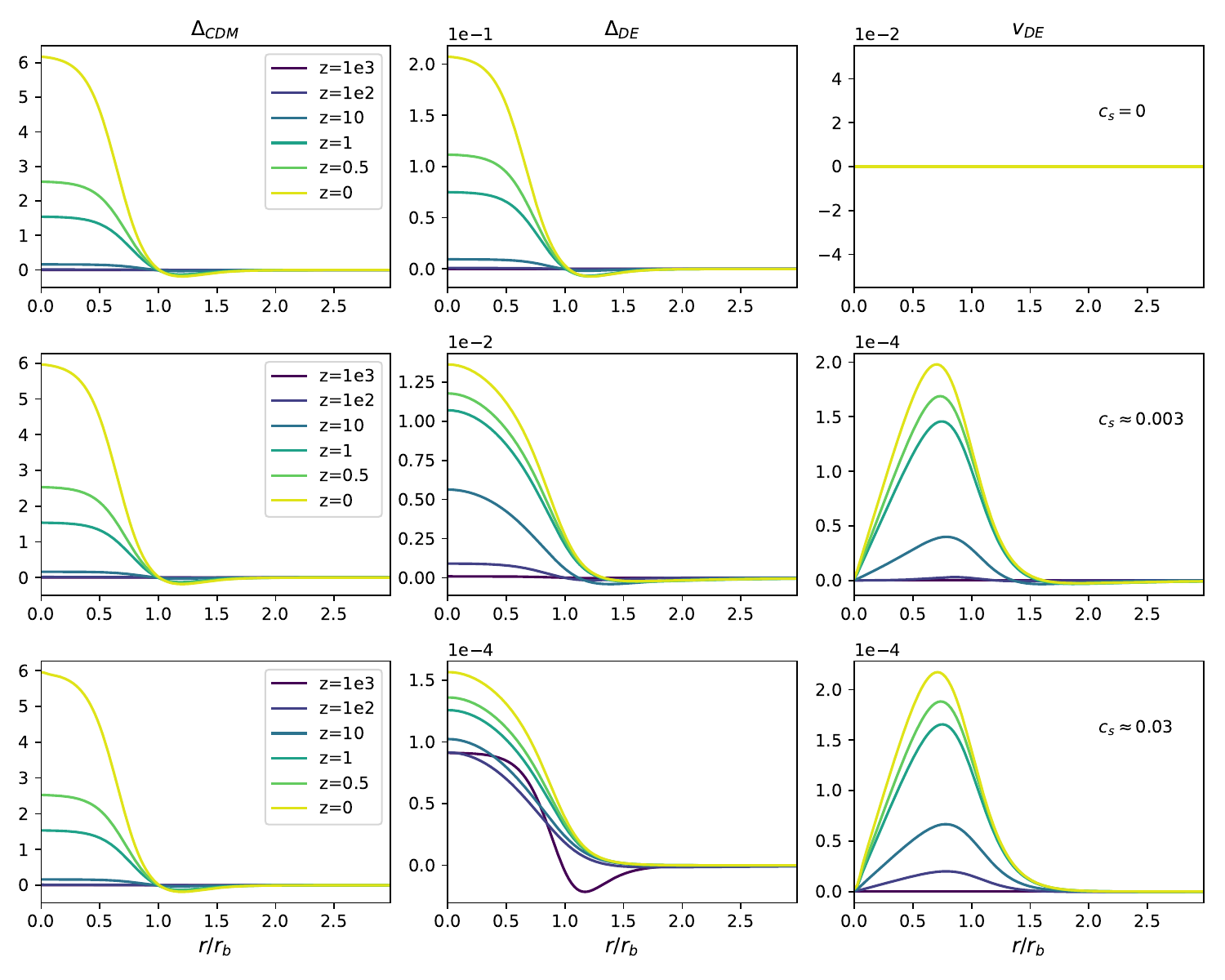}
    \caption{The time evolution of the density contrast of the dust (left column) and the DE fluid (middle column) alongside the DE fluid velocity (right column) in the DE + dust setup described in the text, for three different sound speeds. Curves correspond to redshifts $\bar z=1000$ (blue), 100, 10, 1, 0.5 and 0 (yellow). The uppermost panels have zero sound speed, {\rm i.e.}~the inhomogeneity stays always super sound horizon. In middle panels, where ${\hat c}_s\approx 0.003$ ($\beta_{\DE}=-1.15\times 10^{-5}$), the inhomogeneity falls within the sound horizon at around $z=10$. In the lower panels, where ${\hat c}_s\approx 0.03$ ($\beta_{\DE}=-1.15\cdot10^{-3}$), the sound horizon exceeds the inhomogeneity scale almost through most of the simulation.}
\label{fig:DE1}
\end{center}
\end{figure}
%

This two-component system was studied earlier in~\cite{Marra:2011zp} using a different parametrization and restricted to the limit of tiny sound speed ${\hat c}_{s,{\DE}} \ll 1$. As another consistency check for our code, we reproduced their results (shown in figure 7 of~\cite{Marra:2011zp}) for the case ${\hat c}_{s,{\DE}} = 0$ (with $\beta_{\DE}=0$). Here we take $\bar w_{\DE}=-0.8$ and use the $g_F$ from Eq.~\cref{densprof} with  
$r_b = 5$ Mpc and $\Delta r = 0.3r_b$. The dust density contrast is set to $\bar \delta_{\DM}^{\rm in}=1.55\times 10^{-3}$ at the initial redshift $\bar z_{\rm in}\equiv 1000$ and the background is assumed to be flat with present-day dimensionless density parameters $\Omega_{{\rm DM}0}=0.3$ and $\Omega_{{\rm DE}0}=0.7$. For the Hubble parameter we use $h=0.7$. Our results are shown in the upper three panels in Fig.~\cref{fig:DE1}, where the left and middle panels show the DM fluid and DE relative density perturbations and the right panels the peculiar velocity of the DE fluid. The effect of the, in this case, homogeneous pressure is to damp the DE perturbation by an order of magnitude as compared with the dust. Nevertheless, with zero sound speed the DE fluid can cluster efficiently.

Systems with non-negligible sound speeds could not be studied in~\cite{Marra:2011zp} due to numerical instabilities, but in our formulation these issues are in control. In the middle row of Fig.~\cref{fig:DE1} we show the results of a simulation with $\beta_{\DE}=-1.15\times10^{-5}$ keeping other parameters as given above. This corresponds to ${\hat c}_{s,{\DE}}\approx 0.003$, so that our initial condition~\cref{adcond} is set when the DE-sound horizon is smaller than the bubble:
\begin{equation}
r_{\rm sh}^{\rm in} \equiv \int_0^{t_{\rm in}} {\rm d}t \frac{\hat c_{s,{\DE}}}{\bar a} 
\approx  \frac{2{\hat c}_{s,{\DE}}}{H_0\Omega^{1/2}_{\rm DM0}}\frac{1}{\sqrt{1+\bar z_{\rm in}}} \approx 1 {\rm Mpc} \; < d \,.
\end{equation}
Even though the sound speed is still very small, the pressure gradients already strongly suppress the DE clustering, reducing the amplitude of the DE-inhomogeneity by an order of magnitude. The DE fluid is also visibly pushed outwards relative to dust. The density perturbation of the dust on the other hand is only very slightly suppressed compared to the case with zero sound speed.  

For the lowest panels we used $\beta_{\DE}=-1.15\cdot 10^{-3}$, corresponding to ${\hat c}_{s,{\DE}} \approx 0.03$. In this case the initial condition is set just barely outside the sound horizon. The DE-clustering is significantly diminished by the large sound speed, and the early evolution of $\Delta_{\DE}$ shows small oscillations. In contrast, the fluid velocities are only slightly different compared to the ones in the middle panel. This is understandable since the dominant contribution to the velocity evolution arises from the metric perturbations, which are largely unaffected by the small perturbations in the DE fluid. As opposed to the previous two cases, this large sound speed setup is prone to numerical instabilities associated with the reflection of sound waves from the grid boundary $r_{gb}$. If one were to evolve this system without a modification to control the reflecting waves, the solution would develop unstable oscillations near the origin, terminating the run before $\bar z = 0$. As we explained in section~\cref{sec:bndrycond}, this instability can be avoided by enforcing the non-reflecting boundary conditions dynamically, by suitably tuned numerical diffusion terms. The practical implementation of the diffusion is discussed in appendix~\cref{appendix:numerical_diffusion}. 

We rerun the code for inhomogeneities of several different sizes and different sets of parameters. The generic outcome was what one could expect: inhomogeneities grow in the DE fluid, and DE can cluster efficiently, but a significant DE density perturbation can survive only if the DE-fluid sound horizon at the end of the run remains smaller than the inhomogeneity:  
\begin{equation}
r_{\rm sh}^{\rm end} \sim \frac{{\hat c}_{s,{\DE}}}{H_0} < r_b \,,
\end{equation}

\paragraph{Cosmological constant limit.}

To finish this section we note that while it naively appears that setting $\beta_{\DE} = 1$ and $\bar w_{\DE} = -1$ in~\cref{weq} gives a cosmological constant, this interpretation only holds at the level of the equation of state. At the level of perturbations the model described by~\cref{weq} is very different. Indeed, at this limit model~\cref{weq} gives ${\hat c}_{s,{\DE}}^2 = -1$, which results in  exponentially growing perturbations. The model also shows divergence of the last term in equation~\cref{eq:finalviequation} for $w_i \to -1$, which is a well-known instability of the Euler equations. Thus, the limit with $\beta_{\DE} = 1$ and $\bar w_{\DE} = -1$ in~\cref{weq} does {\em not} describe a cosmological constant, but another exotic, very unstable fluid. In order to model the cosmological constant, one has to first set $v_i(t_{\rm in},r) \equiv 0$ and $\rho_i(t_{\rm in},r) = \rho_\Lambda$, as well as $w_i'=\dot w_i=0$. This results in $B_i=0$ and also $B_i/(1+w_i) = 0$ in the last term in equation~\cref{eq:finalviequation} for any $w_i$ arbitrarily close to -1. Setting this term to zero also for $w_i=-1$ then guarantees that~\cref{eq:finalviequation,eq:finalrhoiequation} hold $\dot v_i=0$ and $\dot \rho_i=0$ at all times, giving the familiar constant profile for the dark energy density. 

The cosmological constant limit of fluid equations is clearly quite delicate, and we wish to point out that in an inhomogeneous setting even an initially homogeneous DE-fluid with $w_i\neq -1$ will necessarily develop inhomogeneities sourced by the metric perturbations. A DE with a phantom equation of state $w_i < -1$ is particularly sensitive to this issue. Thus, treating phantom DE models assuming a persisting homogeneity is not consistent. A further study of such models could be an interesting application of our tools.

%
\subsection{Dust and radiation}
%

Our third example is a two-fluid case with dust and radiation components. That is, we shall assume that $\bar w_{\DM} = w_{\DM} = {\hat c}_{s{\DM}}^2 = 0$ and $\bar w_r = w_r = {\hat c}_{sr}^2 = 1/3$. This case was numerically implemented in~\cite{Lim:2013rra}, again using very different variables and coordinate system. We mainly use this model as a template to highlight numerical issues associated with fluids with very large sound velocities. It is a simple matter to adapt our code to this problem by use of the generic formulae provided in the preceding section. Let us point out however, that under an additional assumption of flatness, $\bar \Omega_{\rm DM}+\bar\Omega_{\rm R}=1$, the initial conditions for the $E$~\cref{Eapp} and $\KTt$~\cref{FbarEq3} can also be integrated analytically:
\begin{equation}
\epsilon_{\rm in} 
= \frac{E_{\rm in}}{(\bar a\bar Hr)_{\rm in}^2} 
=-\frac{5}{6}\,G\Big(\frac{\bar\Omega^{\rm in}_{\rm R}}{\bar\Omega^{\rm in}_{\rm DM}}\Big) \bar\Omega^{\rm in}_{\rm DM} 
\bar{\delta}_{\DM}^{\rm in} g_F(r)
\end{equation}
where $G(x) \equiv 1/\big(1 - 2 x + 8 x^2 + 16 x^{5/2}(\sqrt{x} - \sqrt{1+x})\big)$, and
\begin{equation}
\frac{(\KTt)_{\rm in}}{\bar H_{\rm in} } \,= \, 1 - \epsilon_{\rm in} 
+ (\bar \Omega^{\rm in}_{\rm DM} + \frac{4}{3}\bar\Omega^{\rm in}_{\rm R})\bar{\delta}_{\rm DM}^{\rm in} g_F(r)\,.
\end{equation}
Note that $G(x) \approx 8x/3$ for very large $x$, so that for very large $\bar z_{\rm in}$ the initial conditions asymptote to those of a pure radiation universe. We find that the large sound speed associated with radiation can significantly increase the computation time. This is a generic feature that affects the performance of the code, so we will address it in more detail and provide solutions that allow speeding up the code significantly.

%
\begin{figure}
    \centering
    \includegraphics[width=\linewidth]{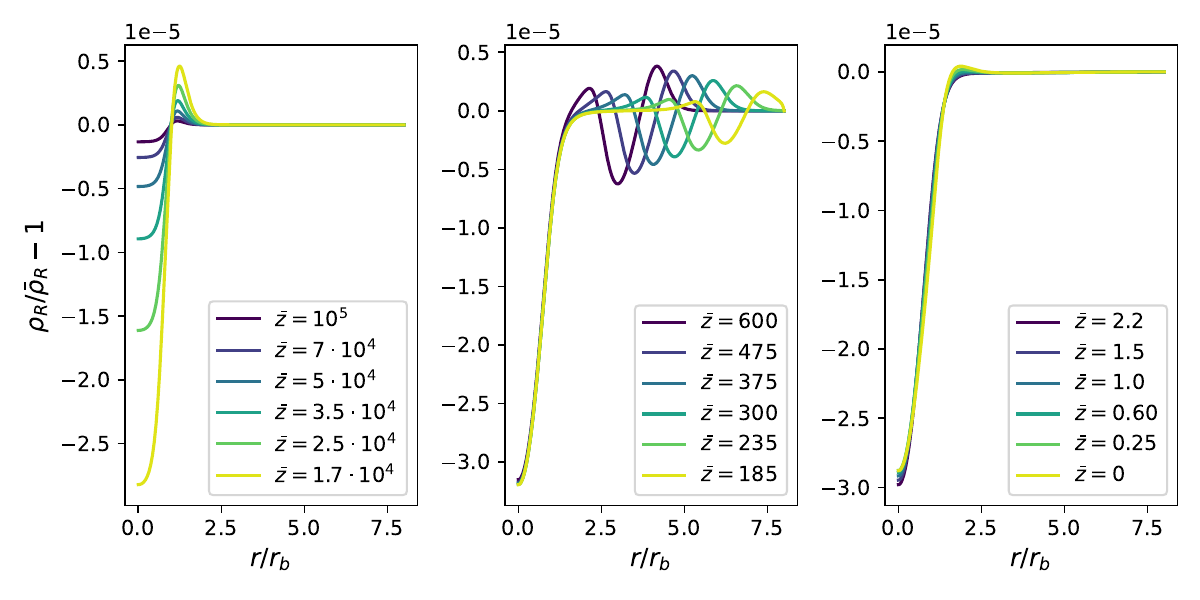}
    \caption{Shown is the evolution of the normalized radiation density perturbation $\rho_\gamma/\bar \rho_\gamma -1$ as a function of the void radius at redshifts falling to three distinct epochs. Left panel shows the early evolution at $\bar z > 1000$ when the density contrast forms. The middle panel shows the evolution at $\bar z\sim 200-600$ when the sound wave hits the grid boundary. The right panel shows the solution at late times, where it displays sharp residual noise on top of the smooth final distribution.}
    \label{fig:radiation_evolution}
\end{figure}
%

In principle, we should start our simulations at very early times for our initial conditions to make sense. This means that all comoving sound horizons should be smaller than the bubble at the initial time:
\begin{equation}
r_{{\rm sh},i}^{\rm in} = \frac{c_{si}}{\sqrt{3}} \int_0^{t_{\rm in}} \frac{{\rm d}t}{\bar a} 
< r_b.
\end{equation}
Neglecting a small correction term arising from the onset of the matter dominance this gives
\begin{equation}
\bar z_{\rm in} > \bar z_{b,si} \approx \frac{c_{si}}{\sqrt{3}H_0\Omega_{\rm R0}^{1/2} r_b},
\label{eq:zinlimit_a}
\end{equation}
where $\bar z_{b,si}$ is the redshift at which the radiation sound horizon in the fluid $i$ reaches the bubble size. This result is valid also for more complex models where the early evolution is governed by radiation. In the $\Lambda$CDM-model in particular, with $\Omega_{\rm R0}\approx 9\times 10^{-5}$ and $H_0 \approx 70$km/s/Mpc we find
\begin{equation}
\bar z_{b,si} \approx c_{si}\, 10^{6} \Big(\frac{{\rm Mpc}}{r_b}\Big) .
\label{eq:zinlimit_b}
\end{equation}
For a structure with $r_b = 10$ Mpc one would thus need $\bar z_{b,si}/c_{si} > 10^5$, while for a Gpc-inhomogeneity $\bar z_{b,si}/c_{si} > 10^3$ would suffice.

The bounds~\cref{eq:zinlimit_a,eq:zinlimit_b} are in fact way too strong in many, if not in most cases. For example, if one is interested in the final nonlinear structures in dark matter and baryons, or, perhaps in the radiation fluid velocity distribution (we will give an example of that below), one may well start the evolution at much smaller redshifts. What happens then is that the initial (unphysical) radiation density profile quickly collapses due to pressure gradients. But unless the early radiation fluctuation had some intrinsic physical value this does not matter, since its effect on other fluids and to the radiation velocity evolution are mainly mediated by the metric perturbations, to which it has hardly any effect after a short initial period. However, even when the radiation density is physically irrelevant its evolution can be highly complicated, involving numerical instabilities and rapid oscillations, which cause the aforementioned problems with the numerical solution. So rather than giving a bound on $\bar z_{\rm in}$, the redshift $\bar z_{b,s\gamma}$ indicates when a more efficient approximation for the radiation density perturbation would be useful.

We illustrate this issue with an example where we take $r_b = 50$ Mpc, $\Omega_{\rm R0}=4.9\times 10^{-5}$ and $\Omega_{\rm DM0}=1 - \Omega_{\rm R 0}$, and start the evolution at a very high redshift $\bar z = 10^5$ that respects the bound~\cref{eq:zinlimit_b}. The initial DM density perturbation was chosen to be $\bar\delta_{\rm DM}^\text{in} = 10^{-6}$. The left panel of figure~\cref{fig:radiation_evolution} shows very early epoch where the density evolves quickly and a sound wave is forming and starting to move outwards. The middle panel displays the time when the sound wave is hitting the computational grid boundary. Most of the perturbation gets later suppressed by numerical diffusion, but some residual always persists and reaches the origin. The right panel shows the late time evolution displaying sharp small scale noise imprinted on top of a smooth density profile. This noise forms when the incoming sound waves hit origin. While these features have negligibly small amplitude, following them numerically can increase the computation time by orders of magnitude. 

\begin{figure}
\begin{center}
\includegraphics[width=0.85\textwidth]{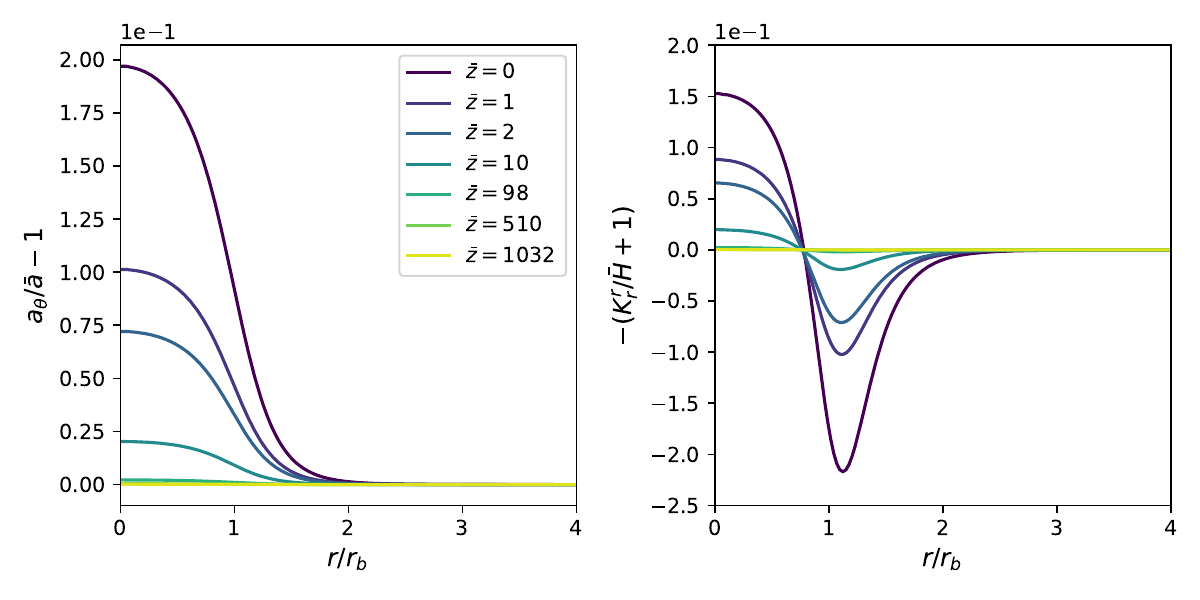}
\caption{Left panel: The evolution of the metric component $a_\theta$ in the setup with dust and radiation with $r_b = 50$ Mpc. Right panel: The evolution of the extrinsic curvature component $K^r_{\phantom{a}r}$ in the same setup, computed using the momentum constraint~\cref{momconst}.}
\label{FigRadiationMetric}
\end{center}
\end{figure}

This spurious noise gets continuously regenerated in radiation when $\bar z \lesssim \bar z_{b,s\gamma}$, and while diffusion clearly controls it to some extent without destroying the overall shape of the radiation density perturbation, it does not fully cure the efficiency problem. However, from the above discussion it is evident that one can simply ignore or outright freeze the radiation density profile for $\bar z \lesssim \bar z_{b,s\gamma}$ with little effect on other fluids and the metric. Indeed, in our example the DM density, metric components (see figure~\cref{FigRadiationMetric}) do not collapse at the onset of $\bar z<\bar z_{b,s\gamma}$, and we have checked that to the numerical accuracy they also remain independent of the evolution of the radiation density profile at $\bar z \lesssim \bar z_{b,s\gamma}$. This discussion may be applied also to other large sound speed fluids when $\bar z \lesssim \bar z_{b,si}$ and so we have implemented the possibility of freezing the distribution in order to conserve computation time into our base code.

%
%

%
\subsection{Ray tracing in the simulated metric}
%

\begin{figure}
\begin{center}
\includegraphics[width=\linewidth]{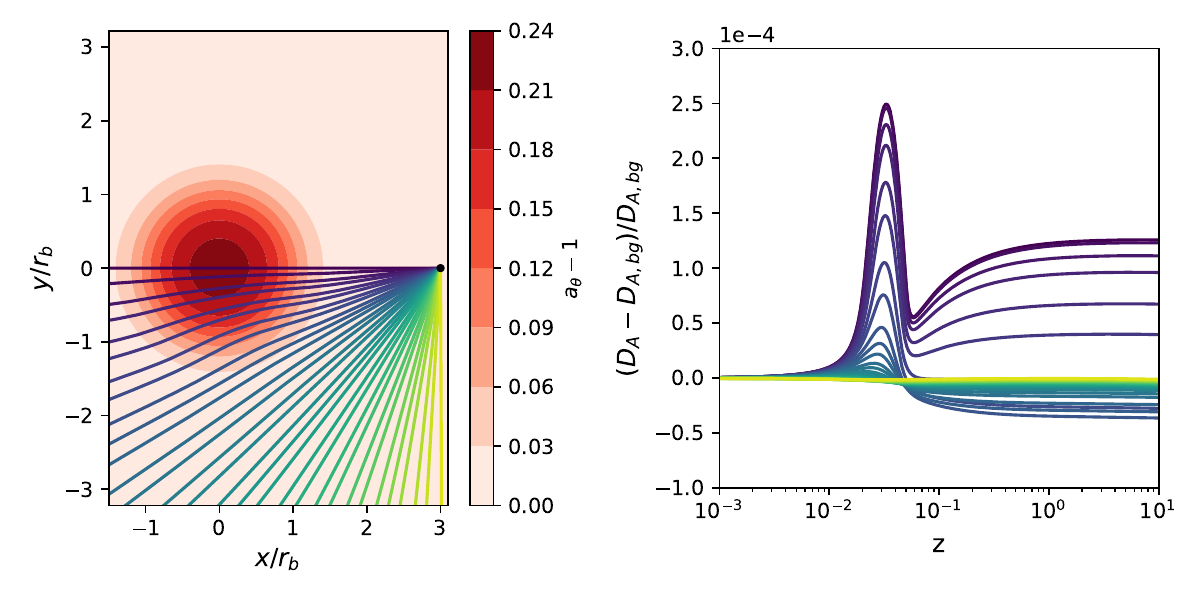} 
\caption{The trajectories of light rays originating from an observer at $r/r_b = 3.0$ and $z_{obs} = 0$ outside an underdensity with $r_b = 50$ Mpc (left panel) and the corresponding perturbations in the angular distances compared to the background FLRW-spacetime measured along the rays as a function of the observed redshifts (right panel). The trajectories are superimposed on the contours of the metric component $a_\theta$ at $z = 0$.}
\label{figAngularDistance}
\end{center}
\end{figure}

Next, we demonstrate the performance of the light ray solver and compute some basic observables using precomputed metric solutions. We used a model with dark matter, radiation and a cosmological constant with $\Omega_{\rm DM0} = 0.3$, $\Omega_{\rm R0} = 4.9\cdot 10^{-5}$, $\Omega_{\Lambda} = 1-\Omega_{\rm DM0} - \Omega_{\rm R0}$ and a void with $r_b=50$ Mpc. The initial dust density perturbation at $\bar z = 10^5$ was chosen to be $\delta^\text{in}_{\DM} = 10^{-6}$ and the rest of the initial perturbations were set according to Eq.~\cref{eq:ballesteros}. In figure~\cref{figAngularDistance} we display the effect of the void to the apparent redshift-angular distance relation, as seen by an observer located at a radius $r_{\text{obs}} = 150$ Mpc outside the void. Rays arriving from different angles are bent and travel through different parts of the void as shown in the left panel of figure~\cref{figAngularDistance}. As a result, the redshift-distance relation acquires the redshift-dependent spread shown in the right panel of figure~\cref{figAngularDistance}.

\begin{figure}[t]
\begin{center}
\includegraphics[width=0.85\textwidth]{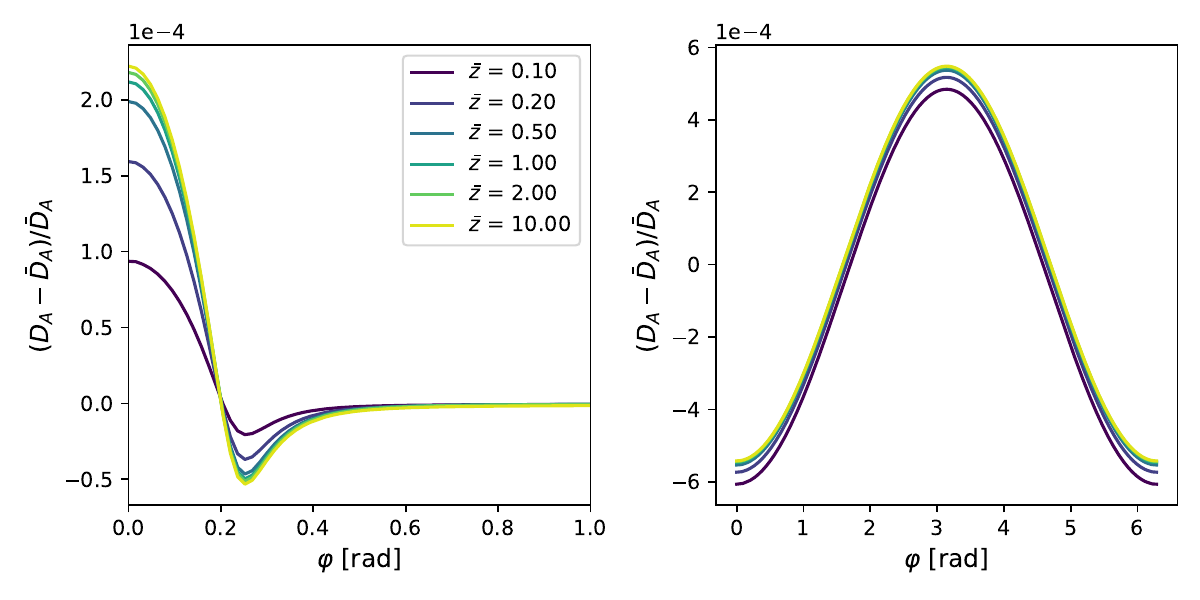} 
\caption{Difference in the observed angular distance from a FRW-solution at different source background redshifts for a void with $r_b = 50$ Mpc, as a function of the observation angle $\phi$. Left panel shows this for an observer well outside a spherical  at $r/r_b = 5$ and the right panel for one inside the void at $r/r_b = 0.5$.}
\label{figAngularDistancediff}
\end{center}
\end{figure}

In the left panel of figure~\cref{figAngularDistancediff} we show the relative deviation from the background value of the angular distance at various redshifts as a function of the observation angle $\phi$, as seen by an observer at $r/r_b = 5$, well outside the void. Orientation is chosen such that $\phi=0$ corresponds to direction through the void center. The right panel shows the same information for an observer inside the void at $r/r_b=0.5$. We see that the amplitude of the difference is much smaller for the outside observer, which is as expected, as the gain in angular distance due to the void is partially canceled by the potential of the void wall~\cite{Marra:2007pm}. On the other hand, the outside observer sees a larger spread as a function of the angle, which is again as expected. Indeed, as long as the observer is deep enough within the void, the angular distance is well fit by a dipole, but the dipole-shape begins to deform as the observer closes in and eventually passes through the void wall. 

%
%

%
%

\begin{figure}
\begin{center}
\includegraphics[width=0.85\textwidth]{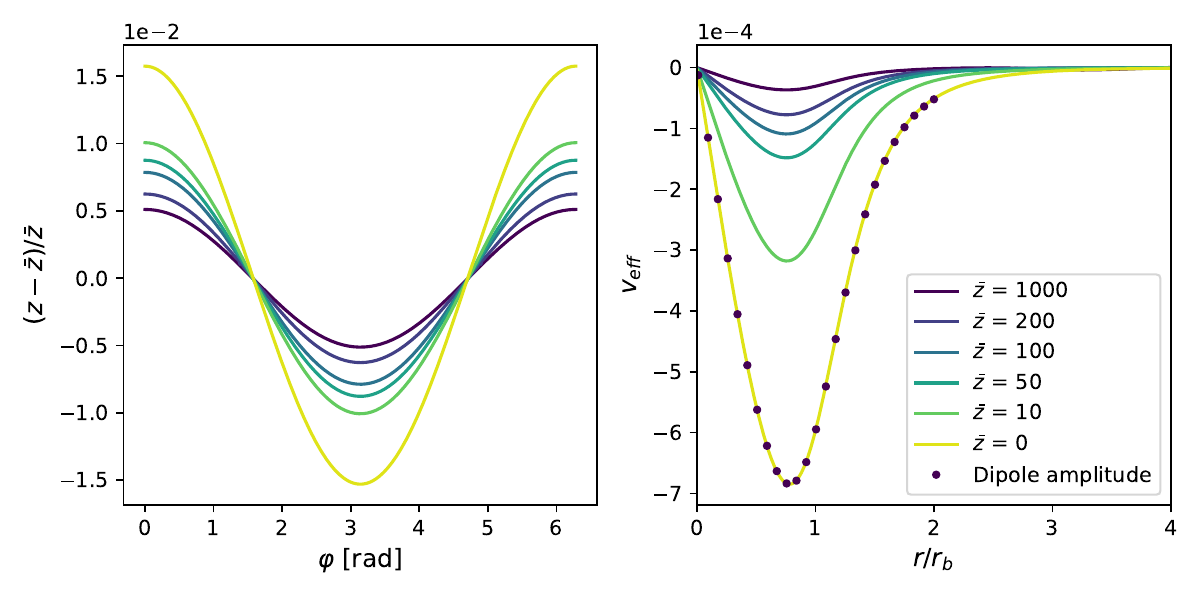}
\caption{Left panel: The relative perturbation in redshift measured by an observer at $r/r_b = 0.5$ at various source background redshifts. The radius of the void is $r_b = 50$ Mpc. Right panel: The asymptotic dipole amplitudes of the radiation temperature $\Delta T = (1+\bar z)/(1+z)-1$ measured by observers located at various radii $r$ in the same setup alongside the radiation fluid peculiar velocity at different time-slices.}
\label{FigKsz}
\end{center}
\end{figure}

Finally, we perform a sanity check for the light ray solver by computing the radial dependence of the effective peculiar velocity involved in a kinematic Sunyaev-Zel'dovich (KSZ) measurement using two different methods. The first one is to simply use the radial velocity of the evolving radiation fluid (its difference from the velocity of non-interacting CDM fluid in the local inertial frame), and the second is to fit a dipole to asymptotic temperature differences $\Delta T = (1+\bar z)/(1+z_\text{obs})-1$ measured by observers at various radii inside the underdensity (see the left panel of figure~\cref{FigKsz} for an example). The right panel of figure~\cref{FigKsz} shows that the two computations agree to a good precision. 

%
%

Let us finally note that in our implementation the light rays should not run directly into the center of the metric. This is just a coordinate singularity due to the spherical coordinate system used, and it can be circumvented by imposing a negligibly small minimum angle. Similarly, a central observer should be displaced from the origin by a small amount. In our testing we found no observable numerical error when rays pass close to the origin. However, one should be aware that insufficient radial resolution of the underlying metric can cause numerical issues in the calculation of the angular diameter distances.

%
\subsection{Interacting dark matter and dark energy}
%

In this subsection we demonstrate how to extend the code to interacting fluids by considering a system of interacting dark matter and dark energy. This model is similar to the model discussed in \cite{cruickshank2025darksectorinteractionsw,Asghari_2019,Simpson_2010}, but our computation is performed in direct space, assuming spherical symmetry and with the full nonlinear evolution. The dark matter component consists of a pressureless dust, while we use the model we defined in section~\cref{sec:de_varying_sound_speed} for the dark energy fluid. Following~\cite{Asghari_2019}, we describe the interaction between the energy components at the fluid level by adding a drag term to the continuity equations:
\begin{equation}
    D_\mu T^{\mu\nu}_{\DM} = +Q^\nu, \qquad\qquad D_\mu T^{\mu\nu}_{\rm D} = -Q^\nu,
\label{eq:DE_DM_coupled_cont_eqs}
\end{equation}
where the drag vector is
\begin{equation}
    Q^\mu = \zeta (u^\mu_{\DE} - u^{\mu}_{\DM})
\label{eq:DMDEcouplingterm}    
\end{equation}
and the positive constant $\zeta$ describes the strength of coupling. In order to implement this model in our code, we first need to reformulate the new continuity equations in the form of equations~\cref{eq:finalviequation,eq:finalrhoiequation}. The equations of motion for energy density and pressure~\cref{eq:rhopequations1} are obtained as suitable linear combinations of components of the continuity equation, and keeping track of the drag term in the calculation results in the modification
\begin{align}
\frac{{\rm d}\rho_\pm}{{\rm d}\tau} &= -\Theta_\pm (\rho_\pm + p_\pm) 
                                  + \gamma_{v_\pm} (Q_\pm^t - v_\pm \ar Q_\pm^r)
\nonumber \\
\frac{{\rm d}p_\pm}{{\rm d}\sigma} &= -a_\pm (\rho_\pm + p_\pm) 
                                 - \gamma_{v_\pm}(v_\pm Q^t_\pm - \ar Q^r_\pm)\,,
\label{eq:modified_rhopequations1}
\end{align}
where $Q^\alpha_\pm \equiv \pm Q^\alpha$ and the sign and index $+$ correspond to dark matter and $-$ to dark energy as in~\cref{eq:DE_DM_coupled_cont_eqs}. These equations can be solved for $\dot \rho_\pm$ and $\dot v_\pm$ as before, which finally yields
\begin{align}
   \dot v_\pm &\!\!\equals\!\! v_\pm \frac{\gamma_{cv_\pm}^2}{\gamma_{v_\pm}^2} 
  \Big( \KRr \!- {\hat c}_{s\pm}^2K
      +   \frac{2{\hat c}_{s\pm}^2v_\pm X}{\rat} - \frac{\gamma^2_{v_\pm}v_\pm'}{\gamma_{c_\pm}^2\ar}\Big)
    - \frac{1}{1\!+\!w_\pm}\frac{B_\pm + I^\pm_{v}}{\gamma_{v_\pm}^2}, 
\label{eq:finalviequationB} \\
   \frac{\dot \rho_\pm}{\rho_\pm} &\!\!\equals \!\!
     -(1\!+\!w_\pm)\gamma_{cv_\pm}^2 \Big( - K 
     + v_\pm^2 \KRr + \frac{2v_\pm X}{\rat} + \frac{v_\pm'}{\ar}  \Big)
     - \frac{v_\pm}{\ar}\frac{\rho_\pm'}{\rho_\pm} + v_\pm B_\pm + I^\pm_{\rho},
\label{eq:finalrhoiequationB}
\end{align}
where the only modification is the presence of the new interaction terms given by 
\begin{align}
  I^\pm_v    &\equiv \frac{\gamma_{cv_\pm}^2}{\rho_\pm}
              \Big( (1+{\hat c}_{s\pm}^2) v_\pm Q^t_\pm 
                  - (1+{\hat c}_{s\pm}^2v_\pm^2) \ar Q^r_\pm \Big)
\nonumber \\  
  I^\pm_\rho &\equiv \frac{\gamma_{cv_\pm}^2}{\rho_\pm}
                 \Big( (1+v_\pm^2) Q^t_\pm - 2v_\pm \ar Q^r_\pm \Big) .
\label{eq:general_Ivrhoterms}
\end{align}
Equations~\cref{eq:finalviequationB,eq:finalrhoiequationB,eq:general_Ivrhoterms} are a fully general set for any two fluids coupled by arbitrary spherically symmetric drag vectors $Q^\mu_\pm = \pm Q^\mu$. Given the simple drag term~\cref{eq:DMDEcouplingterm} one readily finds
\begin{align}
    I^\pm_{v} &\equiv \pm \zeta \frac{\gamma_{cv_\pm}^2}{\rho_\pm} 
    \Big( v_\pm(1 + {\hat c}_{s\pm}^2) (\gamma^{}_{\DE} - \gamma^{}_{\DM}) 
        - (1 + {\hat c}_{s\pm}^2 v_i^2)(\gamma^{}_{\DE}v^{}_{\DE} 
                             - \gamma^{}_{\DM}v^{}_{\DM}) \Big)\,, 
\nonumber \\
    I^\pm_\rho &\equiv \pm \zeta \frac{\gamma_{cv_\pm}^2}{\rho_\pm} 
    \Big ( (1 + v_\pm^2)(\gamma^2_{\DE} - \gamma^2_{\DM}) 
                       - 2 v_\pm(\gamma^{}_{\DE}v^{}_{\DE} 
                             - \gamma^{}_{\DM}v^{}_{\DM}) \Big)\,,
\end{align}
where again the index plus refers to dark matter and minus to dark energy. 

\paragraph{Velocity locking.}

Let us consider the dark energy fluid in the limit $|1 + w_{\DE}| \ll 1$, while still assuming $w_{\DE} > -1$. The effect of the drag is proportional to $1/(1 + w_i)$, which leads to an uneven effect between the fluids. While the effect of the drag on the dark matter can be limited, depending on the choice of $\zeta$, the drag term, along with $B_{\rm DE}$, quickly start to dominate the equations of motion for the dark energy as $w_{\DE} \rightarrow-1$. In this limit, the drag will always lock the velocity of the dark energy fluid to that of the dark matter up to a threshold velocity difference allowed by $B_{\rm DE}$. If the dark energy equation of state has no explicit temporal and radial dependence, $B_{\rm DE}$ is simply a logarithmic gradient term. This leads to the large-scale velocity locking discussed in \cite{cruickshank2025darksectorinteractionsw,Asghari_2019}, whereas inhomogeneities on smaller scales can overcome the drag and lead to more complicated behavior. We can give an order of magnitude estimate for the allowed velocity difference by demanding $|B_{\rm DE}| \sim |I_{v,\DE}|$, neglecting derivatives of $w_{\DE}$ and linearizing the interaction term with respect to velocities, which results in
\begin{equation}
  \Delta v_\text{thr} \sim \frac{\hat c^2_{\DE} |\rho'_{\DE}|}{\zeta \ar}\,.
\label{eq:om-estimate}
\end{equation}
It is worth noting that $B_{\rm DE}=0$ in the homogeneous limit and the drag will always be the dominant term. As such, any motion of the dark matter fluid will lead to the dark energy fluid being dragged along, which necessarily leads to inhomogeneities in the dark energy density of an interacting system. However, the magnitude of this effect will depend on the precise setup.

%
\begin{figure}[t]
    \begin{center}
    \includegraphics[width=0.9\linewidth]{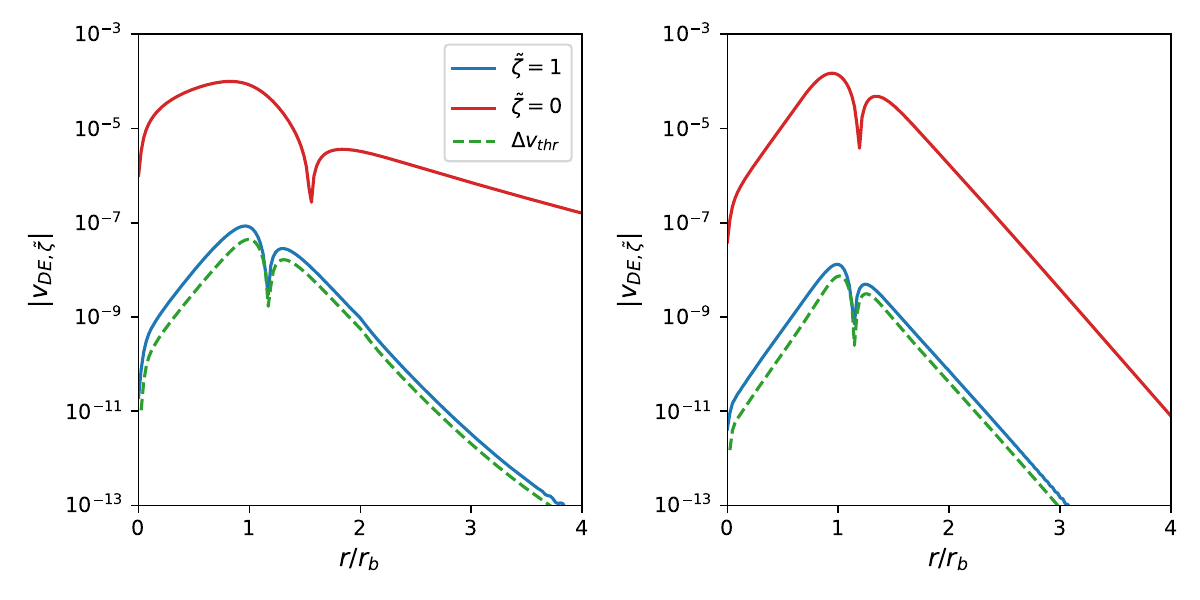}
    \caption{The absolute value of the peculiar velocity of the dark energy fluid with and without drag in the rest frame of the pressureless dust at $\bar z = 0$ for an overdensity with $r_b = 5$ Mpc (left panel) and $r_b = 50$ Mpc (right panel). The dashed line denotes an order of magnitude estimate~\cref{eq:om-estimate} for the allowed velocity difference between dark energy and dark matter, computed using the known dark energy density gradient from the actual solution.}
    \label{drag_vs_nodrag}
    \end{center}
\end{figure}
%

A similar setup has been previously studied in \cite{cruickshank2025darksectorinteractionsw,Asghari_2019,Simpson_2010} using linear perturbation theory, and it was found that even a small velocity difference between the interacting fluids can suppress perturbations in the dark matter fluid. In particular, this effect would be seen on small scales, where the velocity locking does not entirely destroy the velocity differences. Testing how this effect translates beyond linear perturbation theory could provide an interesting avenue for further research. It is also worth noting that for the phantom equation of state $w_{\DE} < -1$ the instability of the fluid equations becomes worse with interaction terms. In our tests, any interaction between fluids led to an unstable growth of perturbations for the phantom dark energy, as long as the initial conditions were not entirely homogeneous for {\em all} fluids in the system. An effective equation of state at $w_{\DE} < -1$, which appears to be preferred by the DESI measurements \cite{DESI:2025fii,DESI:2025zgx,DESI:2025zpo}, is typically achieved with an interacting system of multiple fluids or fields~\cite{M.B.Gavela_2009,Das_Corisanti_Khoury_2006,Guedezounme_2026,Chakraborty_2025}. The behavior of such systems, and the evaluation of how well they can be described by a simple fluid model in the presence of inhomogeneities, could prove an interesting case to study as well. 

%
\subsection{Another example of interacting fluids, baryon-photon coupling}
%

As our last example we show how one can incorporate the usual baryon-photon interaction in our setup. Of course, the density contrasts typically remain very small during the period when the baryon and photon fluids are coupled, and a linear perturbation theory would usually be a more appropriate approach, possibly apart from some exotic models~\cite{Erickcek:2008jp,Erickcek:2008sm}, where the initial radiation density contrast is of order unity at very early times. We then provide the relevant formulae for the coupling, but do not give any numerical examples here. As in the previous section, one introduces a drag term into the photon and baryon continuity equations: 
$D_\mu T^{\mu\nu}_{b} = +Q^\nu$ and   $D_\mu T^{\mu\nu}_{\gamma} = -Q^\nu$, where~\cite{Padmanabhan:1994nc,Baldi:2014ica}:
\begin{equation}
    Q^\mu = - \Gamma \, (h_b)^\mu_{\;\;\nu} (T_\gamma)^\nu_{\;\;\alpha} u^\alpha_b.
    \label{eq:DMDEcouplingterm_2}    
\end{equation}
Here $(T_\gamma)^\nu_{\;\;\alpha} = p_\gamma {g^\mu}_\alpha + (\rho_\gamma + p_\gamma) u_\gamma^\mu u_{\gamma\alpha}$ is the photon energy momentum tensor and the projection operator $(h_b)^\mu_{\;\;\alpha} \equiv {g^\mu}_\alpha + u_b^\mu u_{b\alpha}$ ensures that the drag is orthogonal to baryon velocity: $u_b\cdot Q =  0$. The photon drag on baryons is mediated by free electrons, whose distribution is glued to baryons by Coulomb interactions, so that $\Gamma = X_e n_e \sigma_T$, where $n_e$ is the number density of electrons, $X_e$ is the electron ionization fraction and $\sigma_T$ is the Thomson scattering cross section for electrons. Performing the contractions we find
\begin{equation}
    Q^\mu = -\zeta_{b\gamma} \, (u_\gamma^\mu + (u_\gamma\cdot u_b) u_b^\mu),
    \qquad {\rm with} \qquad \zeta_{b\gamma} = \frac{4}{3}\rho_\gamma \Gamma.
\label{eq:DMDEcouplingterm_3}    
\end{equation}
Using $u_i^\mu =\gamma_i(1;v_i/\ar,0,0)$ one can readily read off the general expressions for $Q^t$ and $Q^r$:
\begin{align}
Q^t &= \zeta_{b\gamma}\gamma_\gamma\Big[ 1- \gamma_b^2(1-\frac{v_bv_\gamma}{\ar^2}) \Big] 
\equiv \zeta_{b\gamma} q^t_{b\gamma}
\nonumber\\
\ar Q^r &= \zeta_{b\gamma}\gamma_\gamma\Big[ v_\gamma 
                - \gamma_b^2(1-\frac{v_bv_\gamma}{\ar^2})v_b \Big]
                \equiv \zeta_{b\gamma} q^r_{b\gamma}.
\label{eq:baryon_photon_drag_general}    
\end{align}
For small velocities, the interaction four-vector is purely spatial to leading order:
\begin{equation}
    Q^t \approx 0, \quad {\rm and} \quad \ar Q^r \approx \zeta_{\gamma b}(v_b-v_\gamma).
\label{eq:baryon_photon_drag_smallv}    
\end{equation}
This implies that at small velocities coupling is only mediated via velocity equations:
\begin{equation}
    I_{\rho,\pm} \approx 0
    \qquad {\rm and} \qquad 
    I_{v,\pm} \approx  \pm \frac{\zeta_{b\gamma} }{\rho_\pm}(v_b-v_\gamma),
\end{equation}
where + (-) refers to baryons (photons). These results are in fact identical with the small velocity limit of the simple interaction terms one used in the interacting DE-DM example. One possible application for this coupling could be in the context of interacting dark matter and dark radiation models~\cite{Schmaltz:2026kgs}, invoked to address the possible slowing down of the linear growth rate relative to $\Lambda \rm CDM$ model at small redshifts $z\lsim 1$~\cite{Macaulay:2013swa,Nguyen:2023fip}.

%
\subsection{Arbitrary number of binary and multi-fluid couplings}
%

Finally, we point out that is simple to generalize our equations to the case of arbitrary number of binary or multi-fluid couplings. In the end, such a system can be described by a number of individual drag terms $Q^\alpha_i$, which are subject to the constraint
\begin{equation}
\sum_i Q_i^\mu = 0.
\label{eq:generalQconstraint}
\end{equation}
The interaction terms in~\cref{eq:general_Ivrhoterms} are already of the correct form for this, and so we only need to add the following interaction terms to all relevant velocity and density contrast equations:
\begin{align}
  I^i_v    &\equiv \frac{\gamma_{cv_i}^2}{\rho_i}
              \Big( (1+{\hat c}_{si}^2) v_i Q^t_i 
                  - (1+{\hat c}_{si}^2v_i^2) \ar Q^r_i \Big)
\nonumber \\  
  I^i_\rho &\equiv \frac{\gamma_{cv_i}^2}{\rho_i}
                 \Big( (1+v_i^2) Q^t_i - 2v_i \ar Q^r_i \Big).
\label{eq:general_Ivrhoterms_i}
\end{align}
where the sum of the drag-vectors $Q_i^\alpha$ is subject to constraint~\cref{eq:generalQconstraint}. These interaction terms are implemented in our base code with optional, user defined drag-vectors. However, if no forces are present, the code defaults to $Q^\alpha_i \equiv 0$. Note that a given drag term $Q_i^\alpha$ can be a sum of different drag vectors, for example arising from binary couplings that the fluid $i$ might have to a number of other fluids in the system.

%
\section{Conclusions and outlook}
%

The limited applicability of the specialized LTB codes and the high complexity of the general-purpose numerical relativity frameworks motivate the development of a solver with a low entry barrier for cosmologists who are not experts in numerical relativity. For these reasons, we have presented LazyTB, a publicly available code for computing cosmological observables in spherically symmetric spacetimes with general energy components. It includes a solver for the fully nonlinear metric, diagnostics for numerical accuracy, and routines to compute observables such as angular and luminosity distances by tracing light rays for observers at arbitrary positions. The spherical ADM equations and general fluid evolution are written in a specific set of dimensionless variables to ensure numerical stability. LazyTB supports an arbitrary number of energy components with user-defined equations of state as well as arbitrary fluid-interactions encoded in terms of optional user defined drag-vectors. The code is written such that it can be straightforwardly extended beyond the cases studied here. LazyTB is available as a Matlab code at \url{https://github.com/ollijvaisanen/LazyTB}, and a Python implementation is under development.

We have demonstrated the performance of the code through a series of working examples, which highlight its simplicity of use, numerical accuracy, and robust stability across different physical setups; in the following, we list and briefly describe each of these examples:
\\
\begin{itemize}

    \item As a first example, we perform a direct comparison between the standard LTB solution and the corresponding result obtained with our code for a dust-only scenario, where the inhomogeneity is embedded in a flat Einstein–de Sitter background. The numerical solution shows excellent agreement, achieving an accuracy at the level of one part in $10^3$, with residual errors fully governed by the chosen numerical tolerances and systematically reducible by increasing the target accuracy.

    \item We extend the dust+DE two-fluid framework by adopting a density-dependent polytropic equation of state for the DE component, and find that DE clustering is controlled mainly by its (adiabatic) sound speed. Dark-energy perturbations can grow, and the DE fluid clusters efficiently, provided its sound horizon stays well below the characteristic scale of the inhomogeneity. 

   \item We apply the formalism to a system of pressureless dust and radiation with high sound speed. The radiation pressure creates sound waves, which reflect from the grid boundary unphysically and crash the code shortly after arriving to the origin. We show that the sound waves can be efficiently suppressed by numerical diffusion, or by freezing the collapsed radiation density distribution with no effect to essential physics. After adding a cosmological constant to this example, we also present our ray-tracing code that allows to compute physical observables, such as angular distances and the kSZ radiation temperature dipoles. 

  \item We extend the code to include couplings between fluids by adding drag-vector interactions to the continuity equations, and apply the extension in a system of interacting dark matter and dark energy. We demonstrated the effect of velocity locking for strong coupling between the fluids. 
  Due to velocity locking, the interacting dust and dark energy fluids can be unified to a single DE fluid with a vanishing sound speed. Unified dark-sector models of this kind have been developed for example in~\cite{Bertacca:2008uf, Lim:2010yk, Bertacca:2010ct}. They are typically constructed by introducing a non-canonical scalar field, which however gives rise to an energy-momentum tensor for a perfect fluid.
    
\end{itemize}

We noted that in DE models with $w_{\DE}\! \neq\! -1$, inhomogeneities in the DE fluid are always generated by metric perturbations, unless the {\em entire} system is completely homogeneous. The effect is particularly strong for the phantom dark energy models, which can not be treated under persisting homogeneity assumption. LazyTB could be used to study multi-fluid systems invoked to get an effective phantom equation of state~\cite{M.B.Gavela_2009,Das_Corisanti_Khoury_2006,Guedezounme_2026,Chakraborty_2025}, and to evaluate how well they can be modeled by a simple fluid model in the presence of inhomogeneities. LazyTB setup could be extended also to treat many self interacting dark matter models with minimal changes. Such models have been motivated by many recent observations, such as~\cite{Cheslog:2026xyr,Gilman:2026wul}.

We showed also how to incorporate the photon-baryon coupling via the electron mediated drag term at the very early universe. While this coupling does not have a practical application, except maybe in the context of large nonlinear early time radiation inhomogeneities~\cite{Erickcek:2008jp,Erickcek:2008sm}, a similar coupling may be relevant for interacting dark matter and dark radiation models~\cite{Schmaltz:2026kgs}, used to explain the apparent slowing down of the structure growth rate at small redshifts $z\lsim 1$~\cite{Macaulay:2013swa,Nguyen:2023fip}.

One obvious application for the LazyTB would be to address the various kinds of directional and redshift space anomalies, such as the quasar dipole~\cite{Secrest:2022uvx} or the bulk flows, or try to explain the features seen in the latest BAO measurements~\cite{DESI:2025zgx}, in particular in the context of alternative theories to the $\Lambda$CDM. While the Hubble tension~\cite{CosmoVerseNetwork:2025alb,H0DN:2025lyy}, might not eventually be amenable to a void-explanation~\cite{Camarena_2021,Camarena_2022} (for a recent attempt see~\cite{Banik:2026xdj} however), there are other local large scale anomalies such as the CMB cold spot~\cite{Finelli:2014yha} or the local KBC void~\cite{Keenan:2013mfa}, whose modeling could  benefit from this software. Also, while the statistical significance of the quasar dipole is under discussion~\cite{Bashir:2025fiv}, it might still be signaling a very large local void~\cite{Cai:2022dov}.

LazyTB could also be used to study the spherical collapse across a wide class exotic models, including the interacting DM and DE models mentioned above, the decaying dark matter, dark matter and dark energy couplings, strongly interacting dark matter (in the fluid limit). While fluid-fluid interactions and finite sound velocities limit the validity of the simplest homogeneous top-hat based spherical collapse models, these features are built into LazyTB. LazyTB can then be used to compute the Halo collapse threshold and virialization functions $\delta_{c}(z_c,M)$ and $\Delta_{\rm vir}(z_c,M)$ in these models without further approximations, and including the dependence on the radial profile of the inhomogeneity. For an alternative approach see~\cite{Batista_2023}. One could then quantify the impact of the alternative theories on the halo concentration–mass relation. Moreover, implementing the LazyTB spherical collapse results within the excursion-set formalism (see {\em e.g.}~\cite{Zentner:2006vw} for a review) would allow estimating the modifications to the halo mass function and the void size distribution. 

Observationally, the most direct consequence of the modified collapse dynamics is a change 
in the abundance of galaxy clusters as a function of redshift and mass proxy. The Dark Energy Survey has completed the analysis of its Year 3 data sets with 16,000 optically selected redMaPPer clusters, and the final full six year Wazp catalog contains more than $40,000$ detected clusters up to $z \simeq  1.3$~\cite{DES:2025tna, DES:2025mqm}. A quantitative comparison with these data sets would require propagating the LazyTB collapse solutions into a calibrated halo mass function and subsequently through the survey selection function and the observable mass relation, including weak lensing mass calibration~\cite{DES:2020ahh}, which would require further work. However, the large available data provides substantial statistical power, and the potentially more accurate modeling of the collapse dynamics allowed by LazyTB might be useful.

Finally, beyond the late-time structure formation problem, a natural extension of our LazyTB would be to include an explicit scalar field sector and use the spherical ADM solver to evolve the Lorentzian, post-nucleation dynamics of isolated vacuum bubbles. Because the scalar field sector in general cannot be described by an ideal fluid, this extension would need extending the main solver code. This would be easy to do however, and the extended framework could be used to study for example the gravitational evolution of an anti-de Sitter bubble within a de Sitter background, with applications to vacuum stability and selected aspects of eternal inflation~\cite{PhysRevD.21.3305, Abbott:1985kr, Johnson:2011wt}. 

%
\begin{acknowledgments}
O.V.~acknowledges the support from the Magnus Ehrnrooth foundation and E.D.S.~acknowledges the support from FONDECYT Project N° 1251141 (Agencia Nacional de Investigaci\'on y Desarrollo, Chile).
We thank Valerio Marra for reading the manuscript and for many valuable discussions and suggestions during the completion of this work. 
\end{acknowledgments}
%

%
\appendix
\addcontentsline{toc}{section}{Appendices}
\section*{Appendices}
%

In appendix~\cref{appendix:A} we provide several technical details pertaining to the numerical implementation and in appendix~\cref{appendix:B} we give explicit expressions for the Ricci and Einstein tensors in variables~\cref{eq:new_metric_variables} as well as the energy momentum tensor needed in the derivation of the source terms in the ADM equations in the main text.

%
\section{Numerical implementation}
\label{appendix:A}
%

We begin by providing the dimensionless form for the ADM equations in appendix~\cref{appendix:ADM_eq} followed by the initial and boundary conditions in~\cref{app:bcond} and the scaled interaction terms in~\cref{appendix:interacting_DMandDE}. We then give the explicit form of Sachs optical equations in~\cref{appendix:sachs_equations}. Finally, in appendix~\cref{appendix:numerical_diffusion} we motivate and detail the implementation of the numerical diffusion terms that damp the unphysical reflecting sound waves. 

%
\subsection{ADM-equations}
\label{appendix:ADM_eq}
%

In this appendix we collect and re-express our evolution equation network in the dimensionless form that is used in the numerical code. Here we explicitly use constant lapse gauge $\alpha = 1$, eliminate ${K^r}_r$ from the equation network by use of the momentum constraint~\cref{eq:C2}. We also use the Hamiltonian constraint to analytically eliminate $h_r$ from the equation for $\partial_y h_\theta$, which makes a number of potentially numerically unstable terms cancel exactly. Finally, we introduce new scaled, dimensionless variables
\begin{equation}
\eta_\theta \equiv\at/\bar{a}, \qquad b_{i}\equiv - {K^i}_{i}/\bar{H}
\qquad {\rm and} \qquad \Delta_i \equiv \rho_i/\bar{\rho}_i \,.
\label{eq:sacled_metric_variables}
\end{equation} 
It is also useful to scale the distance and background Hubble rates by the current Hubble expansion rate $H_0$ and use the scaled curvature variable $e$:
\begin{equation}
\bar r \equiv H_0r \,, \qquad  \bar h\equiv \bar H/H_0 \,, \qquad e \equiv E/\bar r^2 \,.
\end{equation}
Note that the function $e$ is related to the profile $\epsilon_{\rm in}$ used in~\cref{tbarEq} by $\epsilon_{\rm in}= e(t_{\rm in})[(aH)_0/(aH)_{\rm in}]^2$. Here and in the code we scale the present-day scale factor as $a_0 = 1$. Finally, we replace time with the logarithm of the background scale factor $y\equiv{\rm log}\bar a$. In these variables, the full set of equations can eventually be written as
\begin{align}
\partial_y \eta_\theta &\;\equals\; 
      (b_\theta-1)\eta_\theta \phantom{\frac{X}{X}}
\nonumber \\[1 ex]
\partial_y e &\equals 
      \frac{3}{2\bar r}\chi_\theta  X \sum_i\bar\Omega_i(1 + w_i)\Delta_i v_i\gamma_{v_i}^2
\nonumber \\
\partial_y b_\theta &\equals 
     -\frac{3}{2}\Big[ b_\theta^2 - b_\theta(1 + \sum_i\bar\Omega_i\bar w_i )  
     + \sum_i \bar\Omega_i (w_i + v_i^2)\gamma_{v_i}^2 \Delta_i \Big]
     + \frac{e}{\chi_\theta^2} 
\nonumber \\[-0.8 ex]
\partial_y v_i &\equals v_i \frac{\gamma_{cv_i}^2}{\gamma_{v_i}^2} 
\Big[- b^{\scriptscriptstyle \rm MC}_r + 3 {\hat c}_{si}^2 b  
      + \frac{2{\hat c}_{si}^2v_i}{\bar r\chi_\theta}X  
      - \frac{\gamma^2_{v_i}v_i'}{\gamma_{ci}^2\chi_r}\Big]
      - \frac{1}{1+w_i}\frac{\tilde B_i+\tilde I^i_v}{\gamma_{v_i}^2}  
\nonumber \\
\partial_y \Delta_{i} &\equals 3\Big[ \bar w_i - w_i + (1 + w_i)\nu_i \Big]\Delta_i - \frac{v_i\Delta_i'}{\chi_r} + (v_i\tilde  B_i + \tilde I^i_\Delta) \Delta_i. \, 
\label{eq:eomstwofluidscv}
\end{align}
Here prime refers to a derivative with respect to scaled radial coordinate $\bar r$ and we defined $\chi_i \equiv \bar a \bar h \eta_i$, with $\eta_r \equiv \ar/\bar a = (\bar r \eta_\theta)_{,\bar r}/X$ and $X = \sqrt{1+2e\bar r^2}$. The sound velocity was defined in~\cref{eq:sound_velocity} and the time dependent relative background abundances are given by
\begin{equation}
    \bar \Omega_i = \frac{\Omega_{i0}}{\sum_{k=1}^n \Omega_{k0}\exp( 3\int_{0}^y (\bar w_k - \bar w_i ) dy)} \xrightarrow[]{\bar w = \text{const}} \frac{\Omega_{i0}}{\sum_{k=1}^n\Omega_{k0}e^{3y(w_k-w_i)}},
\label{eq:OmegaBar_timedep}    
\end{equation}
where $w_i \equiv p_i/\rho_i$ and $\bar w_i \equiv \bar p_i/\bar \rho_i$. Moreover, the scaled source term $\tilde B_i \equiv B_i/\bar H$ is
\begin{equation}
\tilde B_i =  
\gamma_{cv_i}^2 \left(v_i (\partial_y w_i)_{\rho} + \frac{(\partial_{\bar r} w)_\rho}{\chi_r} + 
\frac{{\hat c}_{si}^2}{\gamma_{v_i}^2\chi_r}\frac{\Delta_i'}{\Delta_i}\right)
\label{eq:Fiequationtwofluidscv}
\end{equation}
and $\nu_i$ refers to the following combination
\begin{equation*}
    \nu_i =  1 - b + \frac13\gamma_{cv_i}^2 \Big( v_i^2 (b^{\scriptscriptstyle\rm MC}_r - 3c_{si}^2 b) 
    - \frac{2v_iX}{\bar r\chi_\theta} - \frac{v_i'}{\chi_r} \Big)
\end{equation*}
The scaled local average Hubble rate is 
\begin{equation}
    b=\frac{1}{3}(b^{\scriptscriptstyle \rm MC}_r + 2b_\theta),
\end{equation} 
and the radial function $b^{\scriptscriptstyle \rm MC}_r$ is defined through the Momentum constraint:
\begin{equation}
b_r^{\scriptscriptstyle\rm MC} = b_\theta 
+ \frac{\bar r\eta_\theta}{(\bar r \eta_\theta)'}\Big(b_\theta'
- \frac{3}{2}\chi_r \sum_i\bar\Omega_i(1 + w_i)\Delta_i v_i\gamma_{v_i}^2 \Big) \,.
\end{equation}
Note that in the background limit $b \to 1$ and $\nu_i \to 0$. Finally, the scaled interaction terms in the velocity and momentum equations are given by 
\begin{align}
  \tilde I^i_v  &\equiv \frac{\gamma_{cv_i}^2}{\bar h^3 \bar \Omega_i\Delta_i}
                \Big( (1+{\hat c}_{si}^2)v_i \bar Q^t_i 
                - (1+{\hat c}_{si}^2v_i^2) \ar \bar Q^r_i \Big)
\nonumber \\  
  \tilde I^i_\Delta &\equiv \frac{\gamma_{cv_i}^2}{\bar h^3 \bar \Omega_i\Delta_i} \Big((1+v_i^2)\bar Q^t_i - 2v_i \ar \bar Q^r_i\Big) .
\label{eq:general_Ivrhoterms_scaled_repeat}
\end{align}
where we defined
\begin{equation}
   \bar Q^\alpha_i \equiv \frac{8\pi G_N}{3H_0^3}Q^\alpha_i,
\end{equation}
and the drag-vectors $Q^\alpha_i$ are subject to constraint $\sum_i Q^\alpha_i = 0$. Note that we took background time dependence explicitly into account in~\cref{eq:general_Ivrhoterms_scaled_repeat} so that $\bar Q^\alpha_i$-functions have a constant scaling. In this way, the user does not need to solve the background Friedmann equations separately for their implementation in the code. Finally, we note that the actual code is written in terms of perturbations for the best numerical accuracy in the early parts of the evolution.

%
\subsection{Boundary and initial conditions}
\label{app:bcond}
%

Boundary conditions are, by design, particularly simple in the new variables: 
\begin{equation}
\eta_\theta(t,r_b) = b_\theta(t,r_b) = \Delta_i(t,r_b) = 1 \qquad {\rm and} \qquad e(t,r_b) = v_i(t,r_b) = 0\,.
\label{eq:scaledboundaryconditions}
\end{equation}
In the code we are actually using $\eta_\theta-1$, $b_\theta-1$ and ${\rm log}\,\Delta_i$ as dependent variables, which also vanish at the boundary. Note that here we assume that $w_i \to \bar w_i$ for all fluids at the grid boundary.

The initial conditions for the dimensionless metric components read as follows: 
\begin{equation}
\eta_\theta^{\rm in} = 1, \quad 
e_{\rm in} = -\frac{1}{2} \sum_i \left[ \bar\Omega_{i0} \bar{a}^{-3w_i-1}_{\rm{in}} \delta^{\rm in}_i \bar R^{\rm in}_ig_{F,i}(r)\right] 
\quad {\rm and} \quad  
b^{\rm in}_{\theta} = - \frac{(\KTt)_{\rm in}}{\bar H_{\rm in}}\,,
\end{equation}
where the shape $g_{F}(r)$ is given by Eq.~\cref{densprof} and perturbations $\delta^{\rm in}_i$ were defined in~\cref{eq:ballesteros} and the $R_i$-functions in~\cref{eq:RandIvariables}. Finally, the initial scaled extrinsic curvature $(\KTt)_{\rm in}/\bar H_{\rm in}$ was already defined in~\cref{FbarEq3}. 

The initial conditions for the fluid variables are
\begin{equation}
v_i^{\rm in} = 0 , \qquad {\rm and} \qquad \Delta_i^{\rm in} = 1 + \bar \delta_i^{\rm in}\Big(1+\frac{\bar r}{3}\partial_{\bar r}\Big)
g_F(r)  \,.
\end{equation}

One does not need to include an equation for the dust, because its velocity and density are given analytically by Eqs.~\cref{eq:dustdens,eq:dustdensity} in the $\alpha = 1$ gauge. In the current scaled variables these become:
\begin{equation}
v_{M}(t,r) \equiv 0 \qquad {\rm and} \qquad \Delta_M(t,r) = 
\frac{\Delta_M^{\rm in}}{\sqrt{1+2\bar r^2e_{\rm in}}}\,
\frac{\sqrt{1+2\bar r^2e}}{ \eta_\theta^2(\eta_\theta+\bar r\eta_{\theta_,\bar r})}  \,.
\label{eq:dustdensity2}
\end{equation}
Of course all sums in the source terms above include a dust contribution with $w_M = \bar w_M = 0$ and the velocity and density given by~\cref{eq:dustdensity2}. 

The Hamiltonian constraint equation, which we use to check the numerical accuracy of our solutions in new variables is:
\begin{equation}
 b^{\scriptscriptstyle\rm HC}_r = -\frac{1}{2}b_\theta + \frac{1}{b_\theta}
 \left( \frac{2e+\bar r e_{,\bar r}}{\chi_\theta\chi_r X }
 + \frac{e}{\chi_\theta^2} \right) + \frac{3}{2b_\theta}\sum_i \bar \Omega_i(1+w_iv_i^2)\gamma_{v_i}^2\Delta_i \,.
\label{eq:finalhctrepeatscv}
\end{equation}
Since our $b_i$-variables are scaled to be of order one, numerical accuracy is guaranteed if at all times $b^{\scriptscriptstyle\rm HC}_r- b^{\scriptscriptstyle\rm MC}_r \ll 1$.

%
\subsection{Interacting fluids examples}
\label{appendix:interacting_DMandDE}
%

We finally list the scalings of the interaction terms in our DE-DM coupling and baryon-photon coupling cases. In the former case the scaled drag-vectors appearing in~\cref{eq:general_Ivrhoterms_scaled_repeat} become
\begin{align}
    \bar Q^t_\pm &= \pm \frac{8\pi G_N}{3H_0^3}\zeta (\gamma^{}_{\DE} - \gamma^{}_{\DM}) 
\\
    \bar Q^r_\pm &= \pm \frac{8\pi G_N}{3H_0^3}\zeta (\gamma^{}_{\DE}v^{}_{\rm \DE} - \gamma^{}_{\DM}v^{}_{\DM}\Big).
\end{align}
where the plus signs apply for DM and minus signs to DE. When implementing this particular system it must bear in mind that the code performs its computations in the rest frame of a non-interacting pressureless dust, which differs from the rest frame of the interacting dark matter fluid, that is no longer described by equation~\cref{eq:dustdensity2}. Even when one wants to model a system containing {\em only} interacting dark matter, it is necessary to add the interacting dark matter as {\em new} fluid, and keep the standard CDM with a vanishing density fraction: $\Omega_\text{CDM} = 0$. This CDM fluid then acts just as a bookkeeping device, setting the reference frame in the system.

In our baryon-photon coupling example the general scaled drag-vectors are
\begin{equation}
  \bar Q^\alpha_\pm  
  = \pm \frac{4}{3} \frac{\rho_\gamma X_e n_e \sigma_T}{\rho_{c,0}H_0} q^\alpha_{\gamma b} 
  \approx \pm 0.0018\, \frac{\Omega_{\gamma 0}}{h} e^{-7y}\, X_e(y)\,q^\alpha_{\gamma b},
\end{equation}
where + sign refers to baryons and - sign to photons and $y=\log(a/a_0)$. To get the last form we used $\rho_\gamma/\rho_{c,0} = \bar\Omega_{\gamma,0}e^{-4y}$, $\sigma_T = (8\pi/3)(\alpha/m_e)^2$ and $n_e = n_B/(1+n_n/n_p) \approx (7/8)\eta_B n_\gamma$, with $\eta_B \equiv n_B/n_\gamma \approx 6.04\times10^{-10}$~\cite{ParticleDataGroup2024} and $n_\gamma = (2\zeta(3)/\pi^2)T_{\gamma,0}^3e^{-3y}$. The functions $q^t_{\gamma b}$ and $q^r_{\gamma b}$ were defined in~\cref{eq:baryon_photon_drag_general}. In the small velocity limit $q^t_{\gamma b} \approx 0$ and $q^r_{\gamma b} \approx v_\gamma-v_b$.

%
\subsection{Optical equations}
\label{appendix:sachs_equations}
%

The optical equations can be written explicitly by taking into account the spherical symmetry, specifying the lapse function $\alpha \equiv 1$. We also restrict ourselves to the plane $\theta = \pi/2$. Here we will use the metric 
\begin{equation}
    ds^2 = - dt^2 + \ar^2 dr^2 + (\rat)^2 d\Omega^2.
\end{equation}
As every quantity is scaled with a factor of $H_0$ in the numerical implementation, we omit the tildes here in order to avoid clutter. With these choices the contracted Ricci tensor becomes
\begin{align}
    R_{\mu\nu} k^\mu k^\nu &= (k^0)^2 \left( -\frac{2 \ddat}{\at} - \frac{\ddar}{\ar} \right) 
\nonumber \\
    & + k^0 k^r \left( 
      \frac{4 \atp\dar}{\at \ar} - \frac{4 \dat'}{\at} - \frac{4 \dat}{r \at} 
    + \frac{4 \dar }{r\ar}  \right) 
\nonumber \\
    & + (k^r)^2 \left( \frac{2 \ar \dat \dar}{\at} + \ar \ddar + \frac{2 \atp\arp}{\at \ar} 
    - \frac{2 \atpp}{\at}  - \frac{4 \atp}{r \at} + \frac{2 \arp}{r \ar} \right) 
\nonumber \\
    & + (k^\phi)^2 \left( 
    1 - \frac{\at^2}{\ar^2} 
    - \frac{4 r \at \atp}{\ar^2} 
    + \frac{r \at^2 \arp}{\ar^3} \right.
\nonumber \\
    & \qquad\qquad \left. 
    + r^2 \left( \dat^2 + \at \ddat + \frac{\at \dat \dar}{\ar}
    - \frac{(\atp)^2}{\ar^2} 
    + \frac{\at \atp \arp}{\ar^3} 
    - \frac{\at \atpp}{\ar^2} \right) \right).
\label{eq:Rkk-term}    
\end{align}
The second contraction needed is $R_{\alpha\mu\beta\nu} s_A^\alpha s_B^\beta k^\mu k^\nu - \frac12 \delta_{AB} R_{\mu\nu} k^\mu k^\nu$. After some algebra it turns out that the 12-component vanishes, which allows us to set $\sigma_{12} \equiv 0$. The 11- and 22- components give identical results
\begin{align}
    R_{\alpha\mu\beta\nu} &s_1^\alpha s_1^\beta k^\mu k^\nu - \frac12 R_{\mu\nu} k^\mu k^\nu 
\nonumber \\
    &= - k_r^2 \frac{\ddar}{2 \ar^3} 
     + k_0^2 \frac{\ddar}{2 \ar} 
     + k_\phi^2 \frac{1}{2r^2}\left[ \frac{1}{r^2}\left( \frac{1}{\at^4} 
     - \frac{1}{\at^2 \ar^2} \right)  \right.
\nonumber \\
    &\left. \quad  
    - \frac{\arp}{r\at^2 \ar^3}
    + \frac{\dat^2}{\at^4} 
    - \frac{\dat\dar}{\at^3 \ar} 
    - \frac{\ddat}{\at^3} 
    - \frac{(\atp)^2}{\at^4 \ar^2} 
    - \frac{\atp\arp}{\at^3 \ar^3} 
    + \frac{\atpp}{\at^3 \ar^2} \right].
\label{eq:Riemann-kk}
\end{align}

The initial conditions for the Sachs equations are set by evolving the system for a few timesteps semi-analytically as in the local FLRW-universe. During this phase the evolution for the angular distance $D_A$ and the expansion scalar $\theta$ are set according to equations~\cref{eq:DA_semi_analytic}, where the local expansion rate is computed from the background evolution. The error introduced by this approximation shows itself as a small constant shift in the angular distance, that gets smaller when the semiclassical region is reduced, and which can be addressed by a suitable subtraction. The error does not appear to affect the angular profile of the solutions. 

%
\subsection{Numerical diffusion}
\label{appendix:numerical_diffusion}
%

%
\begin{figure}
    \centering
    \includegraphics[width=0.9\linewidth]{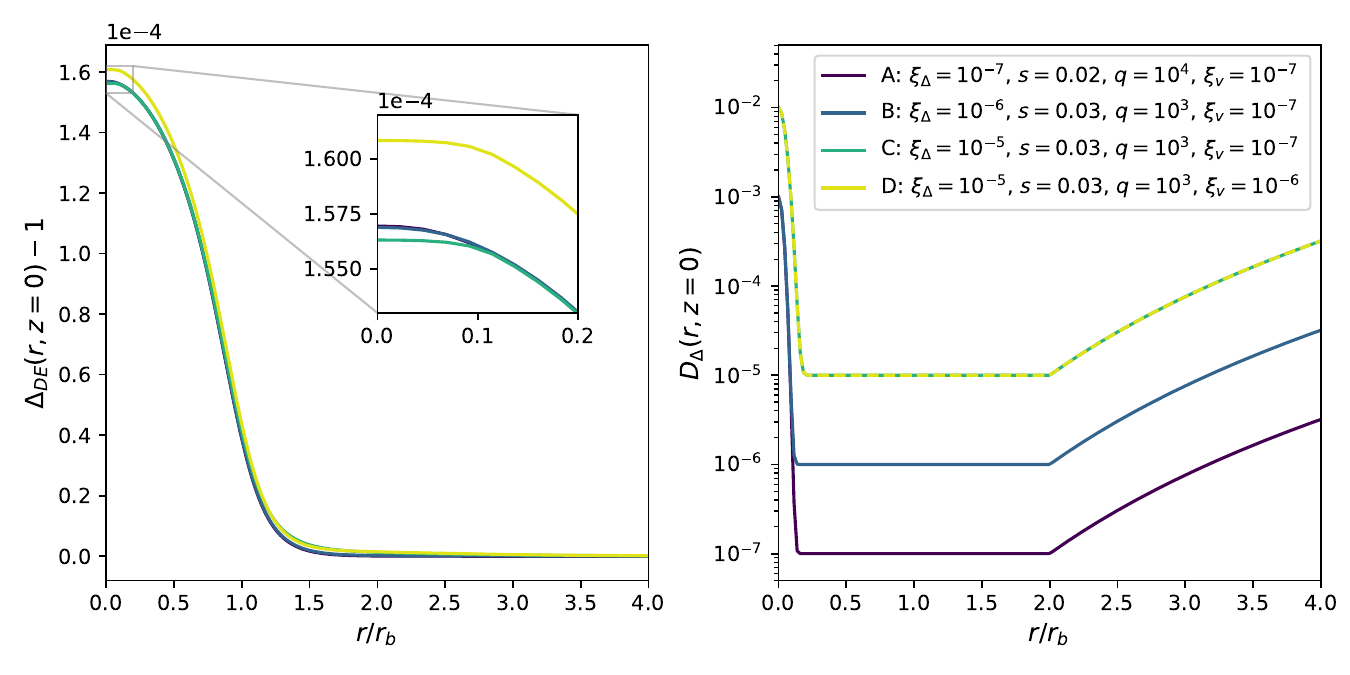}
    \caption{A comparison of density contrast at the end of a run (left panel) computed using various diffusion profiles with diminishing amplitudes (right panel). The diffusion is used to dynamically enforce boundary conditions at the origin and at $r = 8r_b$. All cases were run in the $c_s \approx 0.03$ dynamical DE setup of section~\cref{sec:de_varying_sound_speed}, which was the most sensitive case for diffusion terms discussed in the section.}
    \label{fig:diffusion_comparison}
\end{figure}
%

The evolution equations can become either numerically unstable or prohibitively slow for some initial conditions, particularly if the physical size of the computation grid is small enough compared with the speed of sound in the fluids. A typical instability will manifest itself as growing spurious oscillations originating at $r=0$ and at the outer boundary. One way of mitigating this is to dynamically enforce the boundary conditions (that effectively let all sound waves escape the boundary and allow no incoming waves) by adding numerical diffusion-like terms. The modified equations read
\begin{align}
    \partial_y v_i &= 
    (\partial_y v_i)_{\rm old}
      + \frac{1}{\tilde r^2} \frac{\partial}{\partial \tilde r} 
      \Big[ \tilde r^2 D_v(y,\tilde r) v_i' \Big]
\\
\partial_y \Delta_{i} &= 
(\partial_y \Delta_i)_{\rm old}
+  \frac{1}{\tilde r^2} \frac{\partial}{\partial \tilde r}
    \left[ \tilde r^2 D_\Delta(y,\tilde r)\Delta_i' \right]\,.
\end{align}
where $(\partial_y v_i)_{\rm old}$ and $(\partial_y \Delta_i)_{\rm old}$ refer to all standard terms in the right hand sides of fluid equations, possibly including the interaction terms. In our code, the diffusion coefficients can be freely set by the user, and a good choice generally depends on the exact setup at hand. In our examples, we have used the functions
\begin{align}
    &D_v = \xi_v f(\bar a/\bar a_0) \Big(\frac{\tilde r}{2\tilde r_b}\Big)^5, \label{eq:diffusion_velocity} \\
    &D_\Delta = \xi_\Delta f(\bar a/\bar a_0) \Big[ \theta(2\tilde r_b - \tilde r)(1 + q e^{-(\tilde r/2s \tilde r_b)^2}) + \theta(\tilde r - 2\tilde r_b)\Big(\frac{\tilde r}{2\tilde r_b} \Big)^5 \Big], \label{eq:diffusion_density}
\end{align}
where $f(\bar a/\bar a_0) = 1/3\sqrt{\Omega_{\rm R0} (\bar a/\bar a_0)^4 + \Omega_{\rm DM0}(\bar a/\bar a_0)^3}$ and $\xi_v$ and $\xi_\Delta$ are small constants set individually for each fluid. The profile~\cref{eq:diffusion_density} is depicted on the right panel of figure~\cref{fig:diffusion_comparison} for various parameter choices at $z=0$. 

For the density perturbation, the stability of our flat boundary condition at $r=0$ (von Neumann with zero derivative) is enforced by the narrow gaussian peak in the diffusion function. In our testing such enforcement was not necessary for the velocity perturbation. Similarly, the power law tail damps perturbations reflecting off the outer boundary. Between these two boundaries the diffusion constant is set small enough not to affect the evolution of the perturbation. Diffusion should also be kept small in the early part of the evolution, enforced using the factor $f(a)$, where the sound horizon is still small in comparison with the grid, in order to avoid spoiling the initial conditions. 

The effects of the diffusion terms for different parameter choices in the dynamical DE setup of section~\cref{sec:de_varying_sound_speed}, (with ${\hat c_{sa}} \approx 0.03$), are shown on the left panel of figure~\cref{fig:diffusion_comparison}. The amplitude of the gaussian diffusion peak at $r = 0$ was lowered until the solution near the origin remained visually unchanged (cases C $\to$ B) and in the subsequent profiles the peak amplitude is kept a constant. In this example setup, if the gaussian peak amplitude is lowered further by an order of magnitude, one will begin to observe numerical noise originating from the boundary. Between cases A and B we further narrowed the gaussian peak, with a negligible effect on the solution. However, the peak has to span multiple grid points in order to have the desired effect. Too large diffusion profiles, such as case D, can often be visually identified from the density profile flattening unnaturally. It is also worth noting that despite the small influence on the solution itself, different choices of diffusion terms can have a drastic effect on the computational cost of the runs. For instance, the difference in computation time between cases A and C of figure~\cref{fig:diffusion_comparison} was a factor of two in our testing. The amplitude of the velocity diffusion was similarly lowered (D $\to$ C) until the solution velocity remained unchanged. Suitable diffusion parameters were chosen for the other setups with a similar process.  

%
\section{Tensor Components}
\label{appendix:B}
%

In this appendix we present the nonzero elements of Ricci- and Einstein-tensors for the spherical metric~\cref{sphericmet} in variables~\cref{eq:new_metric_variables}. The non-zero Ricci tensor components are
\begin{equation}
\begin{split}
R_{tt} & = -2\frac{\ddat}{\at} - \frac{\ddar}{\ar} 
\\[-1.5 mm]
R_{tr} & =  2\frac{\ratp\dar}{\rat\ar} - 2\frac{\dratp}{\rat } 
\\[-1 mm]
R_{rr} &=   2\frac{\ratp\arp}{\rat\ar} - 2\frac{\ratpp}{\rat} - 2 \frac{\ar\dat\dar}{\at} + \ar\ddar 
\\[-1.5 mm]
R_{\theta\theta} &=  1 - \Big(\frac{\ratp}{\ar}\Big)^2 + (\drat)^2 + \frac{\rat (\rat)'a'_{r}}{\ar^3} 
                       - \frac{\rat \ratpp}{\ar^2}  + \frac{r^2\at\dat\dar}{\ar} + r^2\at\ddat
\\[0 mm]
R_{\phi\phi} &= \sin^2{\theta}R_{\theta\theta}.
\end{split}
\label{eq:nonzeroricci}
\end{equation}
Ricci scalar is
\begin{equation}
\begin{split}
R & = 2\left( \Big(\frac{\dat}{\at}\Big)^2 + 2\frac{\ddat}{\at} + 2\frac{\dat\dar}{\at\ar} 
                +  \frac{\ddar}{\ar} + \frac{1}{(\rat)^2} 
                + 2\frac{\ratp\arp}{\rat\ar^3} - \Big(\frac{\ratp}{\rat\ar}\Big)^2 
                - 2\frac{\ratpp}{\rat\ar^2}
\right), 
\end{split}
\label{eq:ricciscalar}
\end{equation}
and the non-Zero Einstein tensor components are
\begin{equation}
\begin{split}
G_{tt} & = 2\frac{\ratp\arp}{\rat\ar^3} - \Big(\frac{\ratp}{\rat\ar}\Big)^2 
- 2\frac{\ratpp}{\rat\ar^2} + \frac{1}{(\rat)^2} + \Big(\frac{\dat}{\at}\Big)^2 
+ 2\frac{\dat\dar}{\at\ar} 
\\[-1 mm]
G_{tr} & = 2\frac{\ratp\arp}{\rat\ar} - 2\frac{\dat'}{\at} 
\\[-1 mm]
G_{rr} &=  \Big(\frac{\ratp}{\rat}\Big)^2 - \Big(\frac{\ar}{\rat}\Big)^2 - \Big(\frac{\dat\ar}{\at}\Big)^2 
+ 2\frac{\ar^2\ddat}{\at} 
\\[-1 mm] 
G_{\theta\theta} &= - \frac{\rat\ratp\arp}{\ar^3} + \frac{\rat\ratpp}{\ar^2}  
- r^2\Big( \frac{\at\dat\dar}{\ar} + \at\ddat + \frac{\at^2\ddar}{\ar} \Big) 
\\[0 mm]
G_{\phi\phi} &=  \sin^2{\theta}G_{\theta\theta}.
\end{split}
\label{eq:einsteintensor}
\end{equation}
\vskip0.5cm
\noindent
Finally, we provide the non-zero mixed energy-momentum tensor components
\begin{equation}
\begin{split}
{T^t}_t & = - \sum_i \gamma^2_{v_i}\left(\rho_i + v_i^2p_i\right)  
\\[-2 mm]
{T^t}_r & = \ar \sum_i v_i\gamma^2_{v_i}\left(\rho_i + p_i\right)
\\[-2 mm]
{T^r}_r &=  \sum_i \gamma^2_{v_i}\left(p_i + v_i^2\rho_i\right)  
\\[-2 mm]
{T^\theta}_\theta &= {T^\phi}_\phi = \sum_i p_i .
\end{split}
\label{eq:wLTBemtensor}
\end{equation}
In the pure (dust only) LTB case, only the component ${T^t}_t = -\rho$ is non-vanishing.

%
\bibliographystyle{JHEP}
\bibliography{references}    
%

%
\end{document}